\documentclass{vldb}
\pdfoutput=1

\usepackage{times}
\usepackage{fullpage}
\usepackage{graphicx}
\usepackage{balance} 
\usepackage{color}
\usepackage[utf8]{inputenc}
\usepackage{listings} 
\usepackage{comment}
\usepackage{mdwlist}
\usepackage{subfigure} 
\usepackage{tabularx} 
\usepackage{tikz}
\usepackage{url}
\usepackage{color}
\usepackage{xspace}
\usepackage{appendix}

\newcommand{\subparagraph}{}
\usepackage{titlesec}

\titlespacing{\section}{0ex}{0pt}{0ex}
\titlespacing{\subsection}{0ex}{0.7ex}{0ex}
\titlespacing{\subsubsection}{0ex}{0.7ex}{0ex}
\newcommand{\naive}{na\"{\i}ve\xspace}
\newcommand{\Naive}{Na\"{\i}ve\xspace}

\newcommand{\system}{NAM-DB}
\newcommand*\circled[1]{\tikz[font=\small,baseline=(char.base)]{
            \node[shape=circle,draw,inner sep=1pt] (char) {#1};}}

\newcommand{\Mark}[1]{\textsuperscript#1}

\begin{document}

\title{The End of a Myth: Distributed Transactions Can Scale}
% \numberofauthors{1}
% \author{
% \alignauthor
% \vspace{-8mm}
% \begin{tabular}{cccc}
% Erfan Zamanian\Mark{1}& Carsten Binnig\Mark{1}& Tim Kraska\Mark{1} & Tim Harris\Mark{2}
% \end{tabular}\\
% \vspace{1.5mm}
% \begin{tabular}{cc}
% \affaddr{\Mark{1} Brown University} & \affaddr{\Mark{2} Oracle Labs}\\
% \affaddr{\{erfan\_zamanian\_dolati, carsten\_binnig, tim\_kraska\}@brown.edu} & \affaddr{timothy.l.harris@oracle.com}\\
% \end{tabular}\\
% \vspace{0.75mm}
% }

\numberofauthors{1}
\author{
\alignauthor
%\vspace{-3mm}
\begin{tabular}{cccc}
Erfan Zamanian\Mark{1}\,\,\,& Carsten Binnig\Mark{1}& Tim Kraska\Mark{1} & Tim Harris\Mark{2}\\
\multicolumn{3}{c}{\affaddr{\Mark{1} Brown University}}
& \affaddr{\Mark{2} Oracle Labs}\\
\multicolumn{3}{c}{\affaddr{\{erfan\_zamanian\_dolati, carsten\_binnig, tim\_kraska\}@brown.edu}}
& \affaddr{timothy.l.harris@oracle.com}
\end{tabular}\\
\vspace{0.75mm}
}

%\numberofauthors{2}
% \author{
% \alignauthor
% %\vspace{-8mm}
% Erfan Zamanian\Mark{1} Carsten Binnig\Mark{1} Tim Kraska\Mark{1}
% %\vspace{1.5mm}
% \affaddr{\Mark{1} Brown University}\\
% \affaddr{\{erfan\_zamanian\_dolati, carsten\_binnig, tim\_kraska\}@brown.edu}\\
% %\vspace{0.75mm}
% \alignauthor
% Tim Harris\Mark{2}\\
% \affaddr{Oracle Labs}\\
% \email{timothy.l.harris@oracle.com}
% }

% \numberofauthors{4}
% \author{
% \alignauthor
% \vspace{-8mm}
% Erfan Zamanian\\
% \affaddr{Brown University}\\
% \affaddr{erfanz@cs.brown.edu}
% \alignauthor
% Carsten Binnig\\
% \affaddr{Brown University}\\
% \affaddr{carsten\_binnig@brown.com}
% \alignauthor
% Tim Kraska\\
% \affaddr{Brown University}\\
% \affaddr{tim\_kraska@brown.com}
% \alignauthor
% Tim Harris\\
% \affaddr{Oracle Labs}\\
% \affaddr{timothy.l.harris@oracle.com}
% }

\maketitle

\begin{abstract}

The common wisdom is that distributed transactions do not scale. But what if distributed transactions could be made scalable using the next generation of networks and a redesign of distributed databases? There would be no need for developers anymore to worry about co-partitioning schemes to achieve decent performance. Application development would become easier as data placement would no longer determine how scalable an application is. Hardware provisioning would be simplified as the system administrator can expect a linear scale-out when adding more machines rather than some complex sub-linear function, which is highly application specific. 

In this paper, we present the design of our novel scalable database system \system{} and show that distributed transactions with the very common Snapshot Isolation guarantee can indeed scale using the next generation of RDMA-enabled network technology without any inherent bottlenecks. Our experiments with the TPC-C benchmark show that our system scales linearly to over $6.5$ million new-order ($14.5$ million total) distributed transactions per second on $56$ machines.

%It is commonly (and rightfully) believed that distributed transactions pose major performance bottleneck and significantly hinder the scalability of DBMSs. To avoid distributed transactions as much as possible, DBMSs push the burden of co-partitioning the data onto their users. Sometimes, however, distribution is inevitable due to the nature of the data or workload. In such scenarios, neither DBMSs nor users can do much to avoid poor performance when scaling out.
%What if distributed transactions could be made scalable using the next generation of high-speed networks and a redesign of the distributed DBMS design? There would be no need for developers anymore to worry about co-partitioning schemes to achieve decent performance. Application development would become easier as data placement would no longer determine how scalable an application is. Hardware provisioning would be simplified as the system administrator can expect a linear scale-out when adding more machines rather than some complex sublinear function, which is highly application specific. 

%In this paper, we present the design of our novel scalable distributed database system \system{} and show that distributed transactions with the very common Snapshot Isolation guarantee can indeed scale using the next generation of RDMA-enabled network technology without any inherent bottlenecks. Our experiments with the TPC-C benchmark show that our system scales linearly to over 4.5 million distributed transactions per second on $56$ machines. 
\end{abstract}

%%\vspace{-2.0ex}
\section{Introduction}
\label{sec:intro}

The common wisdom is that distributed transactions do not scale \cite{thecasefordeterministic,h-store,onesize,Schism:VLDB:2010,Andy-thesis,locking12}. 
As a result, many techniques have been proposed to avoid distributed transactions ranging from locality-aware partitioning \cite{Sword:EDBT:2013,HStore:SIGMOD:2012,Schism:VLDB:2010,LocalityAware} and speculative execution \cite{PavloWork} to new consistency levels \cite{consistencyrationing} and the relaxation of durability guarantees \cite{AmrWork}.
Even worse, most of these techniques are not transparent to the developer. 
Instead, the developer not only has to understand all the implications of these techniques, but also must carefully design the application to take advantage of them.
For example, Oracle requires the user to carefully specify the co-location of data using special SQL constructs \cite{OracleCoPartitioning}.
A similar feature was also recently introduced in Azure SQL Server \cite{AzureElasticDB}.
This works well as long as all queries are able to respect the partitioning scheme. 
However, transactions crossing partitions usually observe a much higher abort rate and a relatively unpredictable performance \cite{BinnigDSI}. 
For other application, e.g., social apps, a developer might even be unable to design a proper sharding scheme since those applications are notoriously hard to partition.

But what if distributed transactions could be made scalable using the next generation of networks and redesign the distributed database design?
What if we would treat every transaction as a distributed transaction?
The performance of the system would become more predictable. 
The developer would no longer need to worry about co-partitioning schemes in order to achieve scalability and decent performance. 
The system would scale out linearly when adding more machines rather than sub-linearly because of partitioning effects, making it much easier to provision how much hardware is needed.

Would this make co-partitioning obsolete? Probably not, but its importance would significantly change. 
Instead of being a necessity to achieve a scalable system, it becomes a second class design consideration in order to improve the performance of a few selected queries, similar to how creating an index can help a selected class of queries. 

In this paper, we will show that distributed transactions with the very common Snapshot Isolation scheme \cite{snapshotIsolation} can indeed scale using the next generation of RDMA-enabled networking technology without an inherent bottleneck other than the workload itself (more on that later). 
With Remote-Direct-Memory-Access (RDMA), it is possible to bypass the CPU when transferring data from one machine to another. Moreover, as our previous work \cite{RDMAVision} showed, the current generation of RDMA-capable networks, such as InfiniBand FDR $4\times$, is already able to provide a bandwidth similar to the aggregated memory bandwidth between a CPU socket and its attached RAM. Both of these aspects are key requirements to make distributed transactions truly scalable. 
However, as we will show, the next generation of networks does not automatically yield scalability without redesigning distributed databases. In fact, when keeping the ``old'' architecture, the performance can sometimes even decrease when simply migrating a traditional database from Ethernet network to a high-bandwidth InfiniBand network using protocols such as IP over InfiniBand  \cite{RDMAVision}.

\subsection{Why Distributed Transactions are considered not scalable}

To value the contribution of this paper, it is important to understand why distributed transactions are considered not scalable. 
One of the most cited reasons is the increased contention likelihood. 
However, contention is only a side effect. 
Maybe surprisingly, in \cite{RDMAVision} we showed that the most important factor is the CPU-overhead of the TCP/IP stack.
It is not uncommon that the CPU spends most of the time processing network messages, leaving little room for the actual work.
%while in memory database are getting anyway more and more CPU-bound \cite{oltpThroughGlass}.%\erfan{is this an unfinished sentence? the last part doesn't make much sense}

Additionally, the network bandwidth also significantly limits the transaction throughput. 
Even if transaction messages are relatively small, the aggregated bandwidth required to handle thousands to millions of distributed transactions is high \cite{RDMAVision}, causing the network bandwidth to quickly become a bottleneck, even in small clusters. 
For example, assume a cluster of three servers connected by a 10Gbps Ethernet network. With an average record size of 1KB, and transactions reading and updating three records on all three machines (i.e., one per machine), $6$KB has to be shipped over the network per transaction, resulting in a maximal overall throughput of $\sim 29k$ distributed transactions per second. %(see \cite{RDMAVision} for more details.

%\tim{we should add a back of the envelop calculation here. I had a huge fight with Chris about this. }
Furthermore, because of the high CPU-overhead of the TCP/ IP stack and a limited network bandwidth of typical 1/10Gbps Ethernet networks, distributed transactions have much higher latency, significantly higher than even the message delay between machines, causing the commonly observed high abort rates due to time-outs and the increased contention likelihood; a side-effect rather than the root cause.

Needless to say, there are workloads for which the contention is the primary cause of why distribution transactions are inherently not scalable. 
For example, if every single transaction updates the same item (e.g. incrementing a shared counter), the workload is not scalable simply because of the existence of a single serialization point.
In this case, avoiding the additional network latencies for distributed message processing would help to achieve a higher throughput but not to make the system ultimately scalable.
%which is currently  $\approx 1\mu$s for 1KB over Infiniband FDR compared to $\approx 0.08\mu$s for the CPU to read the same amount of data from memory
%\erfan {I have problem understanding this part. This text reads as if you're saying infiniband FDR will result in higher throughput compared to local CPU?}. 
Fortunately, in many of these ``bottleneck'' situations, the application itself can easily be changed to make it truly scalable \cite{piql,demarcationprotocol}. 
%For instance, most of these contention points can be avoided by making updates commute until an integrity constraint might be violated \cite{consistencyrationing,demarcationprotocol,escrow}.%; i.e., there is a reason why bank transactions usually do not update the balance directly but rather insert a bank-transaction statement. 

\subsection{Why we need a System Redesign}

Assuming the workload is scalable, the next generation of networks remove the two dominant limiting factors to make distributed transaction scale: the network bandwidth and CPU-overhead.  
Yet, it is wrong to assume that the hardware alone solves the problem. 
In order to avoid the CPU message overhead with RDMA, the design of some database components along with their data structures have to change. 
RDMA-enabled networks change the architecture to a hybrid shared-memory and message-passing architecture: it is neither a distributed shared-memory system (as several address spaces exist and there is no cache-coherence protocol) nor is it a pure message-passing system since data can be directly accessed via RDMA reads and writes.
Similarly, the transaction protocols need to be redesigned to avoid inherent bottlenecks. 

While there has been work on leveraging RDMA for distributed transactions, most notably FaRM \cite{farm14,farm15}, most works still heavily rely on locality and more traditional message transfers, whereas we believe locality should be a second class design consideration. 
Even more importantly, the focus of existing works is not on leveraging fast networks to achieve a truly scalable design for distributed databases, which is our main contribution.
Furthermore, our proposed system shows how to leverage RDMA for Snapshot Isolation (SI) guarantees, which is the most common transaction guarantee in practice \cite{EncyclopediaSI} because it allows for long-running read-only queries without expensive read-set validations. 
Other RDMA-based systems focus rather on serializability  \cite{farm15} or do not have transaction support at all \cite{herd}.
At the same time, existing (distributed) SI schemes typically rely on a single global snapshot counter or timestamp; a fundamental issue obstructing scalability. 
%While our findings are based on SI, we believe that they can be generalized to other concurrency schemes as well.

\subsection{Contribution and Outline}
\label{sec:intro:contrib}

In our vision paper \cite{RDMAVision}, we
made the case for how transactional and analytical database systems have to change and showed the potential of taking advantage of efficiently leveraging high-speed networks and RDMA.
In this paper, we follow up on this vision and present and evaluate one of the very first transactional systems for high-speed networks and RDMA. 
In summary, we make following main contributions: 
(1) We present the full design of a truly scalable system called \system{} and propose scalable algorithms specifically for Snapshot Isolation (SI) with (mainly one-sided) RDMA operations. 
In contrast to our initial prototype  \cite{RDMAVision}, the here presented design does no longer restrict the workloads, supports index-based range-request, and efficiently executes long-running read transactions by storing more than one version per record. 
(2) We present a novel RDMA-based and scalable global counter technique which allows to efficiently read the (latest) consistent snapshot in a distributed SI-based protocol.
One of the main limitations of our prototype in \cite{RDMAVision}.
(3) We show that \system{} is truly scalable using a full implementation of the TPC-C benchmark including additional variants where we vary factors such as degree of distribution as well as the contention rate. Most notably, for the standard configuration of the TPC-C benchmark, we show that our system scales linearly to over $3.6$ million  transactions per second on $56$ machines, and $6.5$ million transactions with locality optimizations, that is $2$ million more transactions per second than what FARM \cite{farm15} achieves on $90$ machines. Note, that the total TPC-C throughput would be even higher ($14.5$ million transactions per second) as TPC-C specifies to only report the new-order transactions.

The remainder of the paper is organized as follows:
In Section \ref{sec:architecture} we give an overview of our novel RDMA-based architecture for distributed transaction processing and describe the basic RDMA-based SI protocol as implemented in \cite{RDMAVision}. 
Afterwards, we discuss the main contributions of this paper.
First, in Section \ref{sec:oracle} we explain the design of our new timestamp oracle to generate read and commit timestamps in a scalable manner.
In Section \ref{sec:memory} and Section \ref{sec:compute}, we explain how data is stored in the memory servers and how transactions are executed and persisted.
Afterwards, in Section \ref{sec:evaluation}, we present the results of our experimental evaluation using the TPC-C benchmark.
Finally, we conclude with related work in Section \ref{sec:related} and an outlook on future avenues in Section \ref{sec:concl}.

%\vspace{-2.0ex}
\section{System Overview}
\label{sec:architecture}

In our previous work \cite{RDMAVision}, we showed that the distributed database architecture has to radically change to make the most efficient use of fast networks such as InfiniBand.

InfiniBand offers two network communication stacks: IP over InfiniBand (IPoIB) and remote direct memory access \linebreak (RDMA).
IPoIB implements a classic TCP/IP stack over InfiniBand, allowing existing database systems to run on fast networks without any modifications.
As with Ethernet-based networks, data is copied into OS buffers and the kernel processes them by sending packets over the network, resulting in high CPU overhead and therefore high latencies. 
While IPoIB provides an easy migration path from Ethernet to InfiniBand, IPoIB cannot fully leverage the network's capabilities \cite{RDMAVision}.
On the other hand, RDMA provides a \textit{verbs} API, which enables remote data transfer using the processing capabilities of an RDMA NIC (RNIC).
When using RDMA verbs, most of the processing is executed by the RNIC without OS involvement, which is essential for achieving low latencies.
The verbs API offers two classes of operations: one-sided verbs (read, write and atomic operations) where only the CPU of the initiating node is actively involved in the communication, and two-sided verbs (send and receive) where the CPUs of both nodes are involved. Readers interested in learning more about RDMA are encouraged to read Section 2 in \cite{RDMAVision}. Redesigning distributed databases to efficiently make use of RDMA is a key challenge that we tackle in this paper.

In this section, we first give a brief overview of the network-attached-memory (NAM) architecture proposed in \cite{RDMAVision} that was designed to efficiently make use of RDMA. We then discuss the core design principles of \system{}, which builds upon NAM, to enable a scalable transactional system without an inherent bottleneck other than the workload itself.

\subsection{The NAM Architecture}
\label{sec:architecture:nam}
\begin{figure}
\centering
\includegraphics[width=0.3\textwidth]{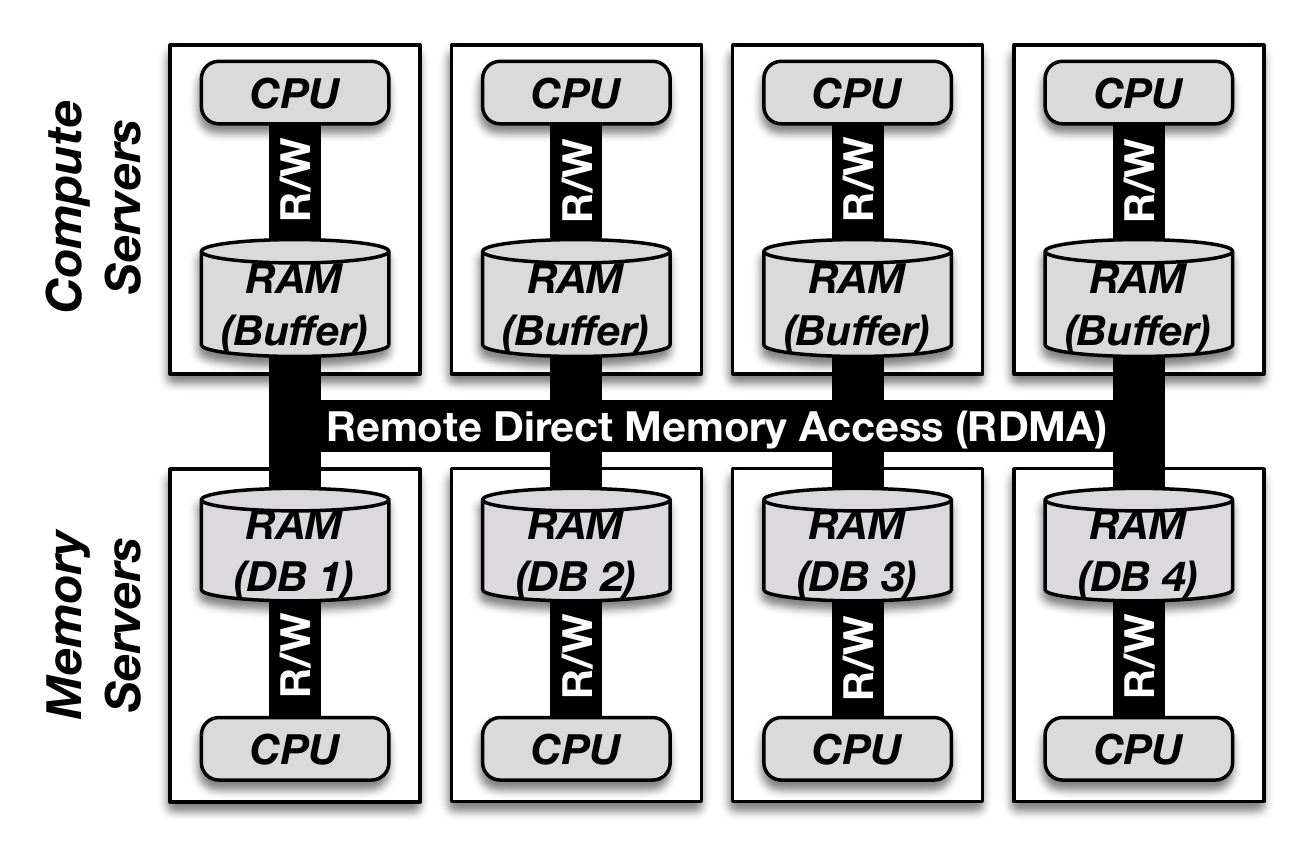}
\vspace{-3.5ex}
\caption{The NAM Architecture}
\vspace{-4.5ex}
\label{fig:namdb}
\end{figure}

The NAM architecture logically decouples compute and storage nodes and uses RDMA for communication between all nodes as shown in Figure \ref{fig:namdb}. 
The idea is that memory servers provide a shared distributed memory pool that holds all the data, which can be accessed via RDMA from compute servers that execute transactions. 
This design already highlights that locality is a tuning parameter. 
In contrast to traditional architectures which physically co-locate the transaction execution with the storage location from the beginning as much as possible, the NAM architecture separates them. 
As a result all transactions are by default distributed transactions. 
However, we allow users to add locality as an optimization like an index, as we will explain in Section~\ref{sec:compute}.
In the following, we give an overview of the tasks of memory servers and compute servers in a NAM architecture. \\

\vspace{-2.5ex}
\textbf{Memory Servers:} In a NAM architecture memory servers hold all data of a database system such as tables, indexes as well as all other state for transaction execution (e.g., logs and metadata).
From the transaction execution perspective, memory servers  are ``dumb'' since they provide only memory capacity to compute servers.
However, memory servers still have important tasks such as memory management to handle remote memory allocation calls from compute servers as well as garbage collection to ensure that enough free space is always available for compute servers, e.g. to insert new records.
Durability of the data stored by memory servers is achieved in a similar way as described in \cite{farm15} by using an uninterruptible power supply (UPS).
When a power failure occurs, memory servers use the UPS to persist a consistent snapshot to disks.
On the other hand, hardware failures are handled through replication as discussed in Section~\ref{sec:compute:recovery}.
\\

\vspace{-2.5ex}
\textbf{Compute Servers:} The main task of compute servers is to execute transactions over the data items stored in the memory servers.
This includes finding the storage location of records on memory servers, inserting/ modifying/ deleting records, as well as committing or aborting transactions.
Moreover, compute servers are in charge of performing other tasks, which are required to ensure that transaction execution fulfills all ACID properties such as logging as well consistency control.
Again, the strict separation of transaction execution in compute servers from managing the transaction state stored in memory servers is what distinguishes our design from traditional distributed database systems. 
As a result, the performance of the system is independent on the location of the data.

\subsection{Design Principles}
\label{sec:architecture:principles}

In the following, we discuss the central design principles of \system{} to enable a truly scalable database system for transactional workloads that builds on the NAM architecture:\\

\vspace{-2.5ex}
\textbf{Separation of Compute and Memory:} 
Although the separation of compute and storage %(i.e., our memory servers) 
is not new \cite{BDS3,deuteronomy15,shared-database:sigmod2015,farm15}, existing database systems that follow this design typically push data access operations into the storage layer.
%However, this design is not ideal if the primary goal is to build a scalable system with reliable performance.
%First, with the recent InfiniBand FDR 4x technology and RDMA the network bandwidth already matches the memory bandwidth of a single machine \cite{RDMAVision}.
%Thus, pushing computation to the data should not be a primary concern when building a scalable distributed main memory databases on fast networks.
%Even more importantly, in modern database systems where all data is located in memory, query processing and transaction execution is increasingly CPU bound \cite{oltpThroughGlass}.
%Consequently, 
However, when scaling out the compute serves and pushing data access operations from multiple compute servers into the same memory server, memory servers are likely to become a bottleneck.
Even worse, with traditional socket-based network operations, every message consumes additional precious CPU cycles. 

In a NAM architecture, we therefore follow a different route. 
Instead of pushing data access operations into the storage layer, the memory servers provide a fine-grained byte-level data access.
In order to avoid any unnecessary CPU cycles for message handling, compute server exploit one-sided RDMA operations as much as possible. % such as RDMA read and write but also atomic operations like RDMA compare-and-swap and fetch-and-add.
This makes the NAM architecture highly scalable since all computation required to execute transactions can be farmed out to compute servers. %\tim{fix this sentence}
%This does not mean that we do not leverage two-sided RDMA operations to push computation into the memory nodes at all.
%However, instead of applying operation push-down as a first class design, we see this just an optimization that can be applied if memory servers are idle anyway.

Finally, for the cases where the aggregated main memory bandwidth is the main bottleneck, this architecture also allows us to increase the bandwidth by scaling out the memory servers.
It is interesting to note that most modern Infiniband switches, such as Mellanox SX6506 108-Port FDR (56Gb/s) ports InfiniBand Switch, are provisioned in a way that they allow the full duplex transfer rate across all machines at the same time and therefore do not limit the scalability.
\\

\vspace{-2.5ex}
\textbf{Data Location Independence:} 
Another design principle is that compute servers in a NAM architecture are able to access any data item independent of its storage location (i.e., on which memory server this item is stored).
As a result, the NAM architecture can easily move data items to new storage locations (as discussed before). Moreover, since every compute server can access any data item, we can also implement work-stealing techniques for distributed load balancing since any compute node can execute a transaction independent of the storage location of the data.

This does not mean that compute servers can not exploit data locality if the compute server and the memory server run on the same physical machine.
However, instead of making data locality the first design consideration in a distributed database architecture, our main goal is to achieve scalability and to avoid bottlenecks in the system design.
Therefore, we argue that data locality is just an optimization that can be added on top of our scalable system. 
\\

\vspace{-2.5ex}
\textbf{Partitionable Data Structures:}
%Finally, the NAM architecture prevents the CPU of a single machine from becoming a bottleneck. 
%Our main principle is that 
As discussed before, in the NAM architecture every compute server should be able to execute any functionality by accessing the externalized state on the memory servers. %(e.g., a global read or commit timestamp).
However, this does not prevent a single memory region (e.g., a global read or commit timestamp) from becoming a bottleneck.%, it helps to mitigate some of the problems and makes scalability issues more transparent from the beginning.
%Our architecture forces the system developer to think about decentralized data structures that can be accessed concurrently and how to efficiently leverage RDMA operations  including atomic operations like fetch-and-add as well as compare-and-swap. 
Therefore, it is important that every data structure is partitionable.
For instance, following this design principle, we invented a new decentralized data structure to implement a partitionable read and commit timestamp as shown in Section \ref{sec:oracle}.

\section{The Basic SI-Protocol}
\label{sec:protocol}

In this section, we describe a first end-to-end Snapshot Isolation protocol using the NAM architecture as already introduced in our vision paper \cite{RDMAVision}.
Afterwards, we analyze potential factors that hinder the scalability of distributed transactions on the NAM architecture, which we then address in  Sections~\ref{sec:oracle}-\ref{sec:compute} as  the main contributions of this paper.

\subsection{A \Naive RDMA Implementation}
\label{sec:protocol:naive}

With Snapshot Isolation (SI), a transaction reads the most recent snapshot of a database that was committed before the begin of the transaction and it does not see concurrent updates. 
In contrast to serializability, a transaction with SI guarantees is only aborted if it wants to update an item which was written since the beginning of the transaction. 
For distributed systems, Generalized SI (GSI) \cite{GeneralizedSI} is more common as it allows any committed snapshot (and not only the most recent one) to be read.
While we also use GSI for \system{}, our goal is still to read a recent committed snapshot to avoid high abort rates. 

\begin{figure}
\centering
\includegraphics[width=0.4\textwidth]{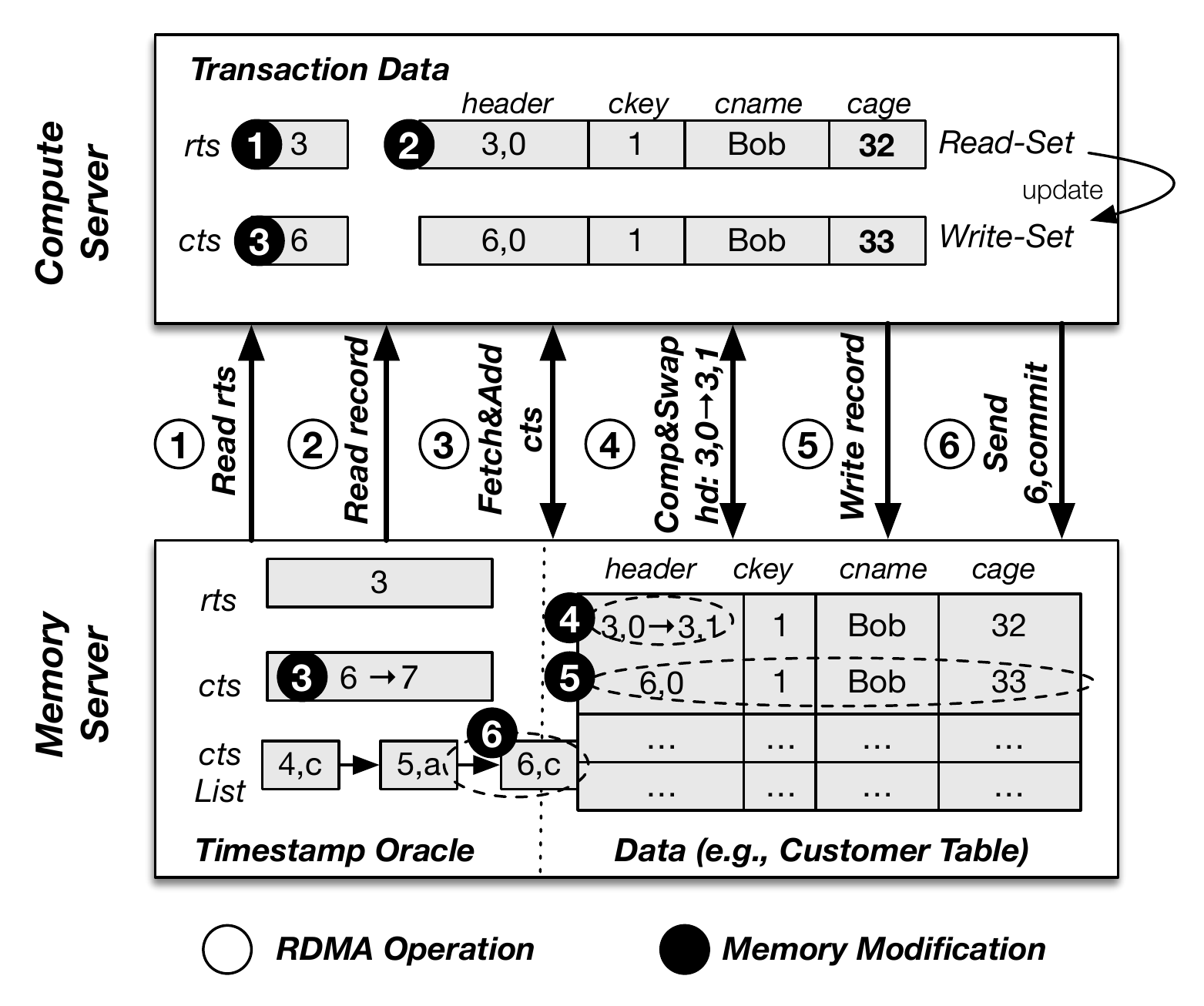}
\vspace{-3.5ex}
\caption{Na\"{\i}ve RDMA-based SI-Protocol}
\vspace{-4.5ex}
\label{fig:rdmasi}
\end{figure}

Listing~\ref{listing:transaction} and the according Figure~\ref{fig:rdmasi} show a \naive{} SI protocol using only one-sided RDMA operations that is based on a global timestamp oracle as implemented in commercial systems \cite{BinnigDSI}. 
The signature of the RDMA operations used in Listing~\ref{listing:transaction} is given in the following table.

\vspace*{0.5ex}
\begin{scriptsize}
\begin{tabularx}{0.95\columnwidth}{|l|X|}\hline
  \textbf{Operation}&\textbf{Signature}\\\hline
  RDMA\_Read & (remote\_addr)\\\hline
  RDMA\_Write & (remote\_addr, value)\\\hline
  RDMA\_Send & (value)\\\hline
  RDMA\_FetchAndAdd & (remote\_addr, increment)\\\hline
  RDMA\_CompAndSwap & (remote\_addr, check\_value, new\_value)\\\hline
\end{tabularx}
\end{scriptsize}
\vspace*{0.5ex}

To better focus on the interaction between compute and memory servers, we made the following simplifying assumptions:
First, we will not consider the durability guarantee and the recovery of transactions.
Second, we assume that there is a catalog service in place, which is always kept up-to-date and helps compute servers find the remote address of a data item in the pool of memory servers;
the remote address of a data item in a memory server is simply returned by the $\&_r$ operator in our pseudocode.
Finally, we consider a simple variant of SI where only one version is kept around for each record and thus we do not need to tackle garbage collection of old versions.
\textbf{Note that these assumptions are only made in this section and we tackle each one later in this paper.}

For \textbf{executing} a transaction, the compute server first fetches the read-timestamp $rts$ using an RDMA read (step \circled{1} in Figure \ref{fig:rdmasi}, line 3 in Listing~\ref{listing:transaction}). 
The $rts$ defines a valid snapshot for the transaction.
Afterwards, the compute server executes the transaction, which means that the required records are read remotely from the memory servers using RDMA read operations (e.g., the record with $ckey=3$ in the example) and updates are applied locally to these records; i.e., the transaction builds its read- and write-set (step \circled{2} in Figure \ref{fig:rdmasi}, line 5 in Listing~\ref{listing:transaction}).
Once the transaction has built its read- and write-set, the compute server starts the commit phase.

For \textbf{committing}, a compute server fetches a unique commit timestamp ($cts$) from the memory server (step \circled{3} in Figure \ref{fig:rdmasi}, line 7 in Listing~\ref{listing:transaction}). 
Fetching a unique $cts$ counter is implemented using an atomic RDMA fetch-and-add operation that returns the current counter and increments it in the memory server by $1$.
Afterwards, the compute server verifies and locks all records in its write-set on the memory servers using one RDMA compare-and-swap operation (line 10-15 in Listing~\ref{listing:transaction}).
The main idea is that each record stores a header that contains a version number and a lock bit in an $8$-Byte memory region.
For example, in Figure \ref{fig:rdmasi}, $(3,0)$ stands for version $3$ and lock-bit $0$ ($0$ means not locked).
The idea of the compare-and-swap operation is that the compute server compares the version in its read-set to the version installed on the memory-server for equality and checks that the lock-bit is set to $0$. If the compare succeeds, the atomic operation swaps the lock bit to $1$ (step \circled{4} in Figure \ref{fig:rdmasi}, line 13 in Listing~\ref{listing:transaction}).

\begin{lstlisting}[belowskip=-10.5ex,mathescape=true,float,floatplacement=H, label=listing:transaction, caption=Transaction Execution in a Compute Server]
runTransaction(Transaction t) {
  // get read timestamp
  rts = RDMA_Read($\&_r$(rts));
  // build write-set
  t.execute(rts);
  // get commit timestamp
  cts = RDMA_FetchAndAdd($\&_r$(cts), 1);

  // verify write version and lock write-set
  commit = true;
  parfor i in size(t.writeSet) {
   header = t.readSet[i].header;
   success[i] = RDMA_CompAndSwap($\&_r$(header), header, setLockBit(header));
   commit = commit &&  success[i];
  }

  // install write-set
  if(commit) {
    parfor i in size(t.writeSet)
      RDMA_Write($\&_r$(t.readSet[i]), t.writeSet[i]);
  }
  //reset lock bits
  else {
    parfor i in size(t.writeSet) {
      if(success[i])
        header = t.readSet[i].header;
        RDMA_Write($\&_r$(header), header); 
    }
  }
  RDMA_Send([cts,commit]); //append cts and result to ctsList
}
\end{lstlisting}

If compare-and-swap succeeds for all records in the write-set, the compute server installs its write-set using RDMA writes (line 19-20 in Listing~\ref{listing:transaction}).
These RDMA writes update the entire record including updating the header, installing the new version and setting the lock-bit back to $0$. For example, $(6,0)$ is remotely written on the header in our example (step \circled{5} in Figure \ref{fig:rdmasi}).
If the transactions fails, the locks are simply reset again using RDMA writes (line 24-28 in Listing~\ref{listing:transaction}).

Finally, the compute server appends the outcome of the transaction (commit or abort) as well as the commit timestamp $cts$ to a list ($ctsList$) in the memory server (step \circled{6} in Figure \ref{fig:rdmasi}, line 32 in Listing~\ref{listing:transaction}).
Appending this information can be implemented in different ways. 
However, for our \naive{} implementation we simply use an unsignaled RDMA send operation; i.e., the compute server does not need to wait for the $cts$ to be actually sent to the server, and give every timestamp a fixed position (i.e., timestamp value - offset) to set a single bit to indicate the success of a transaction.
This is possible, as the fetch and add operation  creates continuous timestamp values. 

Finally, the {\bf timestamp oracle} is responsible to advance the read timestamp by scanning the queue of completed transactions. 
It therefore scans the queue and tries to find the highest commit timestamp (i.e., highest bit) so that every transactions before the timestamp are also committed (i.e., all bits are set). 
Since advancing the read timestamp is not in the critical path, the oracle uses a single thread that continuously scans the memory region to find the highest commit timestamp and also adjust the offset if the servers run out of space. 

\subsection{Open Problems and Challenges}
\label{sec:protocol:problems}
While the previously described protocol achieves some of our goals (e.g., it heavily uses one-sided RDMA to access the memory servers), it is still not scalable. 
The main reason is that global timestamps have inherit scalability issues \cite{abadiwork}, which are emphasized further in a distributed setting. 

First, for every transaction, each compute server uses an RDMA atomic fetch and add operation to the same memory region to get a unique timestamp. 
Obviously, atomic RDMA operations to the same memory location scale poorly with the number of concurrent operations since the network card uses internal latches for the accessed memory locations.
In addition, the oracle is involved in message passing and executes a timestamp management thread that needs to handle the result of each transaction and advance the read timestamp.
Although this overhead is negligible for a few transactions, it shows its negative impact when hundreds of thousands of transactions are running per second.
%Third, RDMA atomic operations require a reliable connection (RC) between compute and memory servers.
%RC's require more state in the network card's memory, resulting in degraded performance when many workers are connected to the same memory server in order to read and write the timestamps.

The second problem with the \naive{} protocol is that it likely results in high abort rates.
A transaction's snapshot can be ``stale'' if the timestamp management thread can not keep up to advance the read timestamp. 
Thus, there could be many committed snapshots which are not included in the snapshot read by a transaction.
The problem is even worse for hot spots, i.e. records which are accessed and modified frequently.

A third problem is that slow workers also contribute to high abort rate by holding back the most recent snapshot from getting updated.
In fact, the oracle only moves forward the read timestamp $rts$ as fast as the slowest worker.
Note that long transactions have the same effect as slow workers.

Finally, the last problem is that the \naive{} implementation does not have any support for fault-tolerance.
In general, fault-tolerance in a NAM architecture is quite different (and arguably less straight-forward) than in the traditional architecture.
The reason is that the transactions' read- and write-sets (including the requested locks) are managed directly by the compute servers.
Therefore, a failure of compute servers could potentially result in undetermined transactions and thus abandoned locks.
Even worse, in our \naive{} implementation, a compute server that fails can lead to ``holes'' in the $ctsList$ causing that the read timestamp does not advance anymore.

Note that beside all these problems, we have made a few simplifying assumptions regarding durability, recovery, meta-data management, version maintenance and garbage collection.
In the remaining sections, we will address these open issues and the problems outlined before.

%\vspace{-2.0ex}
\section{Timestamp Oracle}
\label{sec:oracle}

In this section, we first describe how to tackle the issues outlined in Section \ref{sec:protocol:problems} that hinder the scalability of the timestamp oracle as described in our \naive SI-protocol implementation. 
Afterwards, we discuss some optimizations to further increase the scalability.

%\vspace{-1.0ex}
\subsection{Scalable Timestamp Generation}
\label{sec:oracle:timestamps}

The main issues with the current implementation of the timestamp oracle are: 
(1) The timestamp management thread that runs on a memory server does not scale well with the number of transactions in the system.
(2) Long running transactions/slow compute servers prevent the oracle to advance the read timestamp, further contributing to the problem of too many aborts.
(3) High synchronization costs of RDMA atomic operations when accessing the commit timestamp $cts$ stored in one common memory region. 

In the following, we explain how we tackle these issues to build a scalable timestamp oracle.
The main idea is that we use a data structure called the \textbf{timestamp vector} similar to a vector clock, that represents the read timestamp as the following:

\vspace*{-12pt}
$$T_R = \langle t_1, t_2, t_3, ..., t_n\rangle $$
\vspace*{-12pt}

Here, each component $t_i$ in $T_R$ is a unique counter that is assigned to one transaction execution thread $i$ in a compute server where $i$ is a globally unique identifier.
This vector can either be stored on one of the memory servers or also be partitioned across several servers as explained later. 
However, in contrast to vector clocks, we do not store the full vector with every record but only the timestamp of the compute server who did the latest update:

\vspace*{-20pt}
$$T_C = \langle i, t_i\rangle $$
\vspace*{-12pt}

Here, $i$ is the global transaction execution thread identifier and $t_i$ the corresponding commit timestamp. 
This helps to mitigate one of the most fundamental drawbacks of vector clocks, the high storage overhead per record. \\

\vspace{-2.5ex}
\textbf{Commit Timestamps:} Each component $t_i=T_R[i]$ represents the latest commit timestamp that was used by an execution thread $i$.
Creating a new commit timestamp can be done without communication since one thread $i$ executes transactions in a closed loop. 
The reason is that each thread already knows its latest commit timestamp and just needs to increase it by one to create the next commit timestamp. 
It then uses the previously described protocol to verify if it is safe to install the new versions in the system with timestamp $T_C = \langle i, t+1\rangle $ where $t + 1$ the new timestamp. 

At the end of the transaction, the compute server makes the updates visible by increasing the commit timestamp in the vector $T_R$.
That is, instead of adding the commit timestamp to a queue (line 32 of Listing~\ref{listing:transaction}), it uses an RDMA write to increase its latest timestamp in the vector $T_R$. 
No atomic operations are necessary since each transaction thread $i$ only executes one transaction at a time.  \\

\vspace{-2.5ex}
\textbf{Read Timestamps:} Each transaction thread $i$ reads the complete timestamp vector $T_R$ and uses it as read timestamp $rts$.
Using $T_R$, a transaction can read a record including its header from a memory server and check if the most recent version is visible to the transaction.
The check is simple: as long as the version of the record $\langle i,t\rangle $ is smaller or equal to $t_i$ of the vector $T_R$, the update is visible to the transaction. If not, an older version has to be used so that the condition holds true. 
We will discuss details of the memory layout of our multi-versioning scheme in Section~\ref{sec:memory}.

It is important to note that this simple timestamp technique has several important characteristics.
First, long running transactions, stragglers, or crashed machines do not prevent the read timestamp to advance. 
The transaction threads are independent of each other. 
Second, if the timestamp is stored on a single memory server, it is guaranteed to increase {\bf monotonically}.
The reason is that all RDMA writes are always materialized in the remote memory of the oracle and not cached on its NIC.
Therefore, it is impossible that one transaction execution thread in a compute server sees a timestamp vector like $\langle ..,t_n,.., t_m   +1,..\rangle $ whereas another observes $\langle..,t_n +1,..,t_m,..\rangle $. 
As a result the timestamps are still progressing monotonically, like with a single global timestamp counter. 
However, in the case where the timestamp vector is partitioned, this property might no longer hold true as explained later. 

%One concern is that remote reading $T$ might be relatively expensive since the size of $T$ depends on the number of transaction execution threads that are supported in \system{}.
%For example, for supporting $250$ threads $T$ is $2$KB.
%In order to understand that this is no problem, we refer to our micro-benchmarks in \cite{RDMAVision} which show that latency does not increase tremendously for RDMA read operations up to 2 $KB$. 
%Moreover, in the next section we show techniques to further compress $T$.
%That way, even more transactions threads can be supported without increasing the footprint of $T$.

\subsection{Further Optimizations}
\label{sec:oracle:opti}

In the following, we explain further optimizations to make the timestamp oracle even more scalable:\\

\vspace{-2.5ex}
\textbf{Dedicated Fetch Thread:} Fetching the most recent $T_R$ at the beginning of each transaction can cause a high network load for large transaction execution thread pools on large clusters.
In order to reduce the network load, we can have one dedicated thread per compute server that continuously fetches $T_R$ and all transaction threads simply use the pre-fetched $T_R$.
At a first view this seems to increase the abort rate since pre-fetching increases the staleness of $T_R$.
However, we realized that due to the reduced network load, the runtime of each transaction is heavily reduced, leading to actually a lower abort rate.\\

\vspace{-2.5ex}
\textbf{Compression of $T_R$:} The size of $T_R$ currently depends on the number of transaction execution threads, that could rise up to hundreds or even thousands entries when scaling out.
Thus, instead of having one slot per transaction execution thread, we can compress $T_R$ by having only one slot $t_i$ per compute server; i.e., all transaction execution threads on one machine share one timestamp slot $t_i$.
One alternative is that the threads of a compute server use an atomic RDMA fetch-and-add operation to increase the counter value.
Since the number of transaction execution threads per compute server is bounded (if we use one dedicated thread per core) the contention will not be too high.
As another alternative, we can cache $t_c$ in a compute server's memory. Increasing $t_c$ is then implemented by a local compare-and-swap followed by a subsequent RDMA write.\\

\vspace{-2.5ex}
\textbf{Partitioning of $T_R$:} 
In our evaluation, we found that the two optimizations above are already sufficient to scale to at least $56$ nodes.
However, storing $T_R$ on one memory server could at some point make the network bandwidth of this server the bottleneck.
Fortunately, as a transaction execution thread only needs to update a single slot $t_i$, it is easy to partition $T_R$ across several memory nodes. 
This will improve the bandwidth per server as every machine now only stores a fraction of the vector.
Unfortunately, partitioning $T_R$ no longer guarantees strict monotonicity.
As a result, every transaction execution thread  still observes a monotonically increasing order of updates, but the order of transactions between transaction execution threads  might be different. 
While we believe that this does not impose a big problem in real systems, we are currently investigating if we can solve this by leveraging the message ordering guarantees provided by InfiniBand for certain broadcast operations.
This direction represents an interesting avenue of future work.%, in all our experiments with almost 56 nodes we did not yet observe the need for partitioned timestamp vectors. 

\section{Memory Servers}
\label{sec:memory}

In this section, we first discuss the details of the multi-\linebreak versioning scheme implemented in the memory serves of \system{}, which allows compute servers to install and find a version of a record. 
Afterwards, we present further details about the design of table and index structures in \system|{} as well as memory management including garbage collection.
Note, that our design decision are made to make distributed transactions scalable rather than optimize for locality. 

\subsection{Multi-Versioning Scheme}
\label{sec:memory:layout}

\begin{figure}
	\centering
	\includegraphics[width=0.34\textwidth]{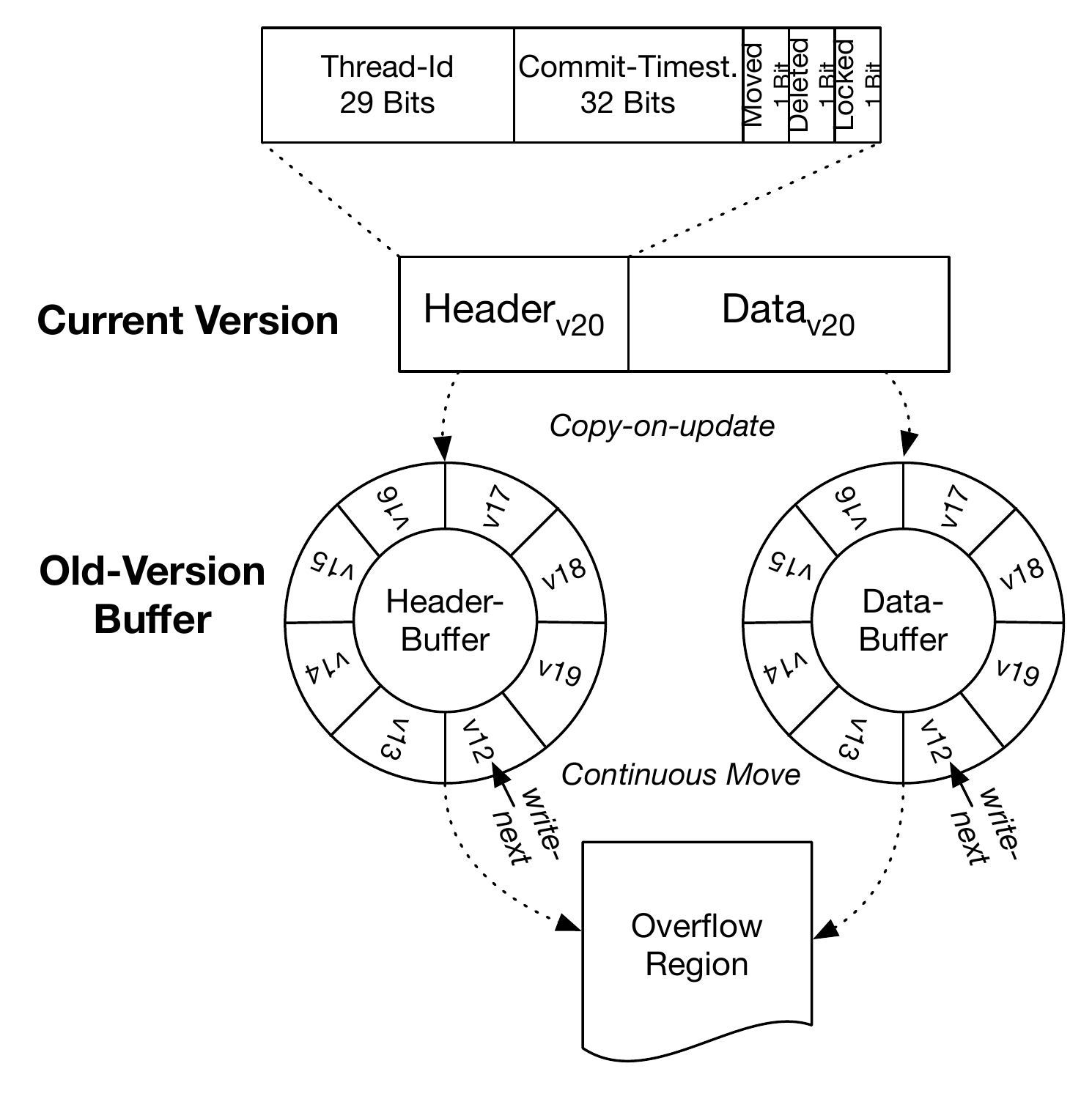}
	\vspace{-3.5ex}
	\caption{Version Management and Record Layout}
	\vspace{-4.5ex}
	\label{fig:layout}
\end{figure}

The scheme to store multiple versions of a database record in a memory server is shown in Figure \ref{fig:layout}.
The main idea is that the most recent version of a record, called the \emph{current version}, is stored in a dedicated memory region.
Whenever a record is updated by a transaction (i.e., a new version needs to be installed), the current version is moved to an \emph{old-version buffer} and the new current version is installed in-place.
As a result, {\bf the most recent version can always be read with a single RDMA request}.
Furthermore, as we use continuous memory regions for the most recent versions, transferring the most recent versions of several records is also only a single RDMA request, which dramatically helps scans. 
The old-version buffer has a fixed size to be efficiently accessible with one RDMA read.
Moreover, the oldest versions in the buffers are continuously copied to an \emph{overflow region}.
That way, slots in the old-version buffer can be re-used for new versions while keeping old versions available for long running transactions.

In the following, we first explain the memory layout in more detail and then discuss the version management.\\
%Finally, we also discuss how installing and reading a version is implemented using one-sided RDMA operations.\\

\vspace{-2.5ex}
\textbf{Record Layout:} For each record, we store a \emph{header} section that contains additional metadata and a \emph{data} section that contains the payload.

The data section is a full copy of the record which represents a particular version.
For the data section, we currently only support fixed-length payloads.
Variable-length payloads could be supported by storing an additional pointer in the data section of a record that refers to the variable-length part, which is stored in a separate memory region.
However, when using RDMA, the latency for messages of up to $2$KB remains constant as shown in our vision paper \cite{RDMAVision}. 
Therefore, for many workloads where the record size does not exceed this limit, it makes sense to store the data in a fixed-length field that has the maximal required length inside a record.

The header section describes the metadata of a record.
In our current implementation we use an $8$-byte value that can be atomically updated by a remote compare-and-swap operation from compute servers.
The header encodes different variables:
The first $29$ bits are used to store the \emph{thread identifier i} of the transaction execution thread that installed the version (as described in the section before). 
The next $32$ bits are used for the \emph{commit timestamp}.
Both these variables represent the version information of a record and are set during the commit phase.
Moreover, we also store other data in the header section that is used for version management each represented by a $1$-bit value: 
a \emph{moved}-bit, a \emph{deleted}-bit, and a \emph{locked}-bit.
The moved-bit indicates if a version was already moved from the old-version buffer to the overflow region and thus its slot can be safely reused.
The deleted-bit indicates if the version of a record is marked for deletion and can be safely garbage collected.
Finally, the locked-bit is used during the commit phase to avoid concurrent updates after the version check.
\\

\vspace{-2.5ex}
\textbf{Version Management:} 
The \emph{old-version buffer} consists of two circular buffers of a fixed size as shown in Figure \ref{fig:layout}; one that holds only headers (excluding the current version) and another one that holds data sections.
The reason for splitting the header and the data section into two circular old-version buffers is that the size of the header section is typically much smaller than the data section.
That way, a transaction that needs to find a particular version of a record only needs to fetch all headers without reading the payload. 
Once the correct version is found, the payload can be fetched with a separate read operation using the offset information.
This effectively minimizes the latency when searching a version of a record.
%More details about how reading and installing versions is implemented using RDMA is explained next.

For installing a new version, we first validate if the current version has not changed since reading it and set the lock-bit using one atomic RDMA compare-and-swap operation (i.e., we combine validation and locking). 
If locking and validation fails, we abort.
Otherwise, the header and the data of the current version is then copied to the old version buffer.
In order to copy the current version to the buffers, a transaction first needs to determine the slot which stores the oldest version in the circular buffer and find out if that slot can be overwritten (i.e., the moved-bit is set to $1$).
In order to identify the slot with the oldest version, the circular buffers store an extra counter which is called the next-write counter.

Another issue is that the circular buffers have only a fixed capacity.
The reason that we efficiently want to access them with one-sided RDMA operations and avoid pointer chasing operations.
However, in order to support long-running transactions or slow compute servers the number of versions stored in the buffer might not be sufficient since they might need to read older versions.
As a solution, a version-mover thread that runs in a memory server continuously moves the header and data section of the old-version buffers to an overflow region and sets the moved-bit in the header-buffer to $1$.
This does not actually mean that the header and data section are removed from the old-versions buffers.
It only means that it can be safely re-used by a transaction that installs a new version.
Keeping the moved versions in the buffer maximizes the number of versions that can be retrieved from the old-version buffers.

\subsection{Table and Index Structures}
\label{sec:memory:table}

In the following we discuss how table and index structures are implemented in memory servers.\\

\vspace{-2.5ex}
\textbf{Table Structures:} In \system{} we only support one type of table structure that implements a hash table design similar to the one in \cite{pilaf}.
In this design, compute servers can execute all operations on the hash table (e.g., put or a get) by using one-sided RDMA operations.
In addition to the normal put and get operations to insert and lookup records of a table, we additionally provide an update to install a new record version as well as delete operation.

The hash tables in \system{} stores key-value pairs where the keys represent the primary keys of the table.
Moreover, the values store all information required by our multi-versioning scheme: the current record version as well as three pointers (two pointers to the old-version buffers as well as one pointer to the overflow region).

Different from \cite{pilaf}, hash tables in \system{} are partitioned to multiple memory servers.
In order to partition the hash table, we split the bucket array into equally-sized ranges and each memory server stores only one of the resulting sub-ranges as well as the corresponding keys and values.
In order to find a memory server which stores the entries for a given key, the compute server only needs to apply the hash function which determines the bucket index and thus the memory server which holds the bucket and its key-value pairs. 
Once the memory server is identified, the hash table operation can be executed on the corresponding memory server.\\

\vspace{-2.5ex}
\textbf{Index Structures:} In addition to the table structure described before, \system{} supports two types of secondary indexes: a hash-index for single-key lookups and a $B^+$-tree for range lookups.
Both types of secondary indexes map a value of the secondary attribute to a primary key that can then be used to lookup the record using the table structure dicussed before (e.g., a customer name to the customer key).
Moreover, secondary indexes do not store any version information.
Retrieving the correct version of a record is implemented by the subsequent lookup on the table structure using the primary key that is returned by the lookup on the secondary index.

In order to implement the hash index in \system{}, we use the same hash table design that we have described before that is used for implementing tables.
The main difference is that values in a secondary hash index, we store only primary keys and no pointers (e.g., to old-version buffers etc.) as discussed before.
For the $B^+$-tree index, we follow a different route.
Instead of designing a tree structure that can be accessed purely by one-sided RDMA operations, we use two-sided RDMA operations to implement the communication between compute and memory server.
The reason is that operations in $B^+$-trees need to chase multiple pointers from the root to the leave level.
When implementing pointer chasing with one-sided RDMA operations, multiple roundtrips between compute and memory servers are required.
This is also true for a linked list in a hash table.
However, as shown in \cite{pilaf} when clustering keys in a linked list into one memory region one RDMA read / write is often sufficient to execute a get / put an entry if no collision occurs.
Moreover, for scaling-out and to avoid that individual memory servers become a bottleneck, we range partition $B^+$-trees  to different memory servers.
In the future, we plan to investigate into alternative indexing designs for $B^+$ trees.% that allows compute servers to leverage one-sided RDMA operations to access them.

\subsection{Memory Management}
\label{sec:memory:management}

Memory servers store tables as well as index structures in their memory as described before.
In order to allow compute servers to access tables and indexes via RDMA, memory servers must pin and register memory regions at the RDMA network interface card (NIC).
However, pinning and registering memory at the NIC are both costly operations which should not be executed in a critical path (e.g., whenever a transaction created a new table or an index).
Therefore, memory servers allocate a large chunk of memory  during initialization and register it to the NIC.
After initialization, memory servers handle allocate and free calls from compute servers.\\

\vspace{-2.5ex}
\textbf{Allocate and Free Calls:} 
Allocate and free calls from compute servers to memory servers are implemented using two-sided RDMA operations.
In order to avoid many small memory allocation calls, compute servers request memory regions from memory servers in extends.
The size of an extend can be defined by a compute server as a parameter and depends on different factors (e.g., expected size and update rate).
For example, when creating a new table in \system{}, a compute server who executed the transaction allocates an extend that allows to store an expected number of records and their different versions.
The number of expected records per table can be defined by applications as a hint.
\\

\vspace{-2.5ex}
\textbf{Garbage Collection:} In order to avoid that tables  constantly grow and need to allocate more and more space from memory servers, old versions of records need to be garbage collected.
The goal of garbage collection is to find out which versions of a record are not required anymore and can be safely evicted.
In \system{}, garbage collection is implemented by having a timeout on the maximal transaction execution time $E$ that can be defined as a parameter by the application.
Transactions that run longer than the maximal execution time might abort since the version they require might already be garbage collected.

In order to implement our garbage collection scheme we capture a snapshot of the timestamp vector $T$ that represents the read timestamp of the timestamp oracle (see Section \ref{sec:oracle}) in regular intervals. 
We currently create a snapshot of $T$ every minute and store it together with the wall-clock time in a list sorted by the wall-clock time.
That way, we can find out which versions can be safely garbage collected based on the maximal transaction execution time.
For garbage collecting these versions, a garbage collection thread runs on every memory server which continuously scans the overflow regions and sets the deleted-bit of the selected versions of a record $1$. 
These versions are then truncated lazily from the overflow regions once contiguous regions can be freed.

\section{Compute Servers}
\label{sec:compute}

In this section, we first discuss how compute servers execute transactions and then present techniques for logging and recovery as well as fault-tolerance.

\subsection{Transaction Execution}
\label{sec:compute:trans}

Compute servers use multiple so called \emph{transaction execution threads} to execute transactions over the data stored in the memory servers.
Each transaction execution thread $i$ executes transactions sequentially using the complete timestamp vector $T$ as read timestamp as well as $(i, T[i])$ as commit timestamp to tag new versions of records as discussed in Section \ref{sec:oracle}.
The general flow of executing a single transaction in a transaction execution thread is the same workflow as outlined already in Section \ref{sec:protocol:naive}.
Indexes are updated within the boundaries of the transaction that also updates the corresponding table using RDMA operations  (i.e., we pay additional network roundtrips to update the indexes).

One import aspect, that we have not discussed so far, is how the database catalog is implemented such that transactions can find the storage location of tables and indexes.
The catalog data is hash-partitioned and stored in memory servers.
All accesses from compute servers are implemented using two-sided RDMA operations since query compilation does not result in a high load on memory servers when compared to the actual transaction execution.
Since the catalog does not change to often, the catalog data is cached by compute servers and refreshed in two cases:
The first case is, if a requested database object is not found in the cached catalog, the compute server requests the required meta data from the memory server.
The second case is, if a database object is altered.
We detect this case by storing a catalog version counter within each memory server that is incremented whenever an object is altered on that server.
Since transaction execution threads run transactions in a closed loop, this counter is read from the memory server that store the metadata for the database objects of a given transaction before compiling the queries of that transaction. 
If the version counter has changed when compared to the cached counter, the catalog entries are refreshed.

\begin{comment}
Finally, we would like to discuss two additional optimizations for implementing the transaction execution in memory servers.
First, in order to minimize latencies when using one-sided RDMA operations it is important to efficiently make use the asynchronous communication model of RDMA where compute servers can register an RDMA read/write operation at the NIC that executes this operation. After registering a read/write operation compute servers can directly continue without waiting for the actual result of this operation and poll for completion later.
We leverage this asynchronous communication model in \system{} heavily to minimize latencies.
Second, another optimization is that we allow compute servers to be physically co-located with memory servers.
In this case, we leverage the locality of data; i.e., transactions use local memory access operations instead of RDMA if possible.
However, it is important to note that atomic RDMA operations and atomic CPU operations are unaware of each other.
Thus, even with our locality optimization, atomic operations are always implemented using RDMA atomics even for local transactions.
\end{comment}

\subsection{Failures and Recovery}
\label{sec:compute:recovery}

\system{} provides a fault-tolerance scheme that tolerates failures of compute and memory servers.
In the following, we discuss both cases.
At the moment, we do not handle failures resulting from network partitioning since the events are extremely rare in InifiniBand networks.
Handling these types of failures, could be handled by using a more complex commit protocol than 2PC (e.g., a version of Paxos based on RDMA) which is an interesting avenue of future work.
Moreover, it is also important to note that a high-availability is not the design goal of \system{}, which could be achieved in \system{} by replicating the write-set during the commit phase.\\

\vspace{-2.5ex}
\textbf{Memory Server Failures:} In order to tolerate memory server failures, each transaction execution thread of a compute server writes a private log journal to a memory server using RDMA writes.
In order to avoid the loss of a log, each transaction execution thread writes its journal to more than one memory server.
The entries of such a log journal are $<T, S>$ where $T$ is the read snapshot used by thread $i$ and $S$ is the executed statement with all its parameters.
Commit timestamps that have been used by a transaction execution thread are stored as parameters together with the commit statement and are used during replay.
The log entries for all transaction statements are written to the database log before installing the write-set on the memory servers.

Once a memory server fails, we halt the complete system
and recover all memory servers to a consistent state from the last persisted checkpoint (discussed below). 
For replaying the distributed log journal, the private logs of all transaction execution threads need to be partially ordered by their logged read timestamps $T$.
Therefore, the current recovery procedure in \system{} is executed by one dedicated compute server that replays the merged log for all memory servers.
%Looking into more efficient designs to recover the memory servers in a NAM-DB architecture is a avenue for future work.

In order to avoid long restart phases, an asynchronous thread additionally writes checkpoints to the disks of memory servers using a dedicated read-timestamp.
This is possible in snapshot isolation without blocking other transactions.
The checkpoints can be used to truncate the log.
\\

\vspace{-2.5ex}
\textbf{Compute Server Failures:} Compute servers are stateless and thus do not need any special handling for recovery.
However, a failed compute server might result in abandoned locks.
Therefore, each compute server is monitored by another compute server called monitoring compute server.
If a monitoring compute server detects that a compute server is down, it unlocks the abandoned locks using the log journals written by the transaction execution threads of this compute server.

\section{Evaluation}
\label{sec:evaluation}

The goal of our experiments is to show that distributed transactions can indeed scale and locality is just an optimization.% like adding an index. 
%In the following, we first discuss the benchmark and setup before we present the key results. 
%In the following, we first discuss the benchmarks and the setup used for this evaluation and then present the results of the individual experiments.

%\subsection{Benchmark and Setup}

\textbf{Benchmark:} As benchmark, we used TPC-C \cite{TPCC}.
We used the standard schema and configuration of TPC-C without any modifications unless otherwise stated.
We generated $50$ warehouses per memory server %(i.e., the database was partitioned by warehouses). We 
and created all required secondary indexes.% to support point- and range-lookups. 
All these indexes were implemented using our hash- and $B^+$-tree index as discussed in Section \ref{sec:memory}.
Moreover, for showing the effect of locality, we added a parameter to TPC-C that allows us to change the degree of distribution for new-order transactions from $0\%$ to $100\%$ (where $10\%$ is the standard configuration).
As defined by the benchmark we only report the throughput of {\em new-order} transactions, which can make up to $45\%$ of the benchmark. 

\textbf{Setup:} For executing the experiments, we used two different clusters both with an InfiniBand network:

\emph{Cluster $A$:} This is a cluster at Oracle that is composed of $57$ machines (two types of machines) in total that are all connected to an InfiniBand FDR 4X network using a Mellanox Connect-IB card.
All machines are attached to the same InfiniBand switch. 
While the first $28$ machines of type 1 provide two Intel Xeon E7-4820 processors (each with 8 cores) and $128$ GB RAM, the other $29$ machines of type 2 provide two Intel Xeon E5-2660 processors (each with 8 cores) and $256$ GB RAM.
All machines in this cluster run Oracle Linux Server 6.5 (kernel 2.6.32)  as operating system and use the Mellanox OFED $2.3.1$ driver for the network.

\emph{Cluster $B$:} This is our own (smaller) InfiniBand cluster with $8$ machines that are all connected to an InfiniBand FDR 4X network using a Mellanox Connect-IB card.
All machines are attached to the same InfiniBand switch.
Moreover, each machine has two Intel Xeon E5-2660 v2 processors (each with 10 cores) and $256$GB RAM.
The machines all run Ubuntu 14.01 Server Edition (kernel 3.13.0-35-generic) as operating system and use the Mellanox OFED $2.3.1$ driver for the network.

For showing the scalability of \system{}, we used Cluster $A$. However, since we only had restricted access to that cluster, we executed the more detailed analysis (e.g., the effect of data locality) on Cluster $B$.

\begin{figure}
\centering
\vspace{-1ex}
\includegraphics[width=0.6\columnwidth]{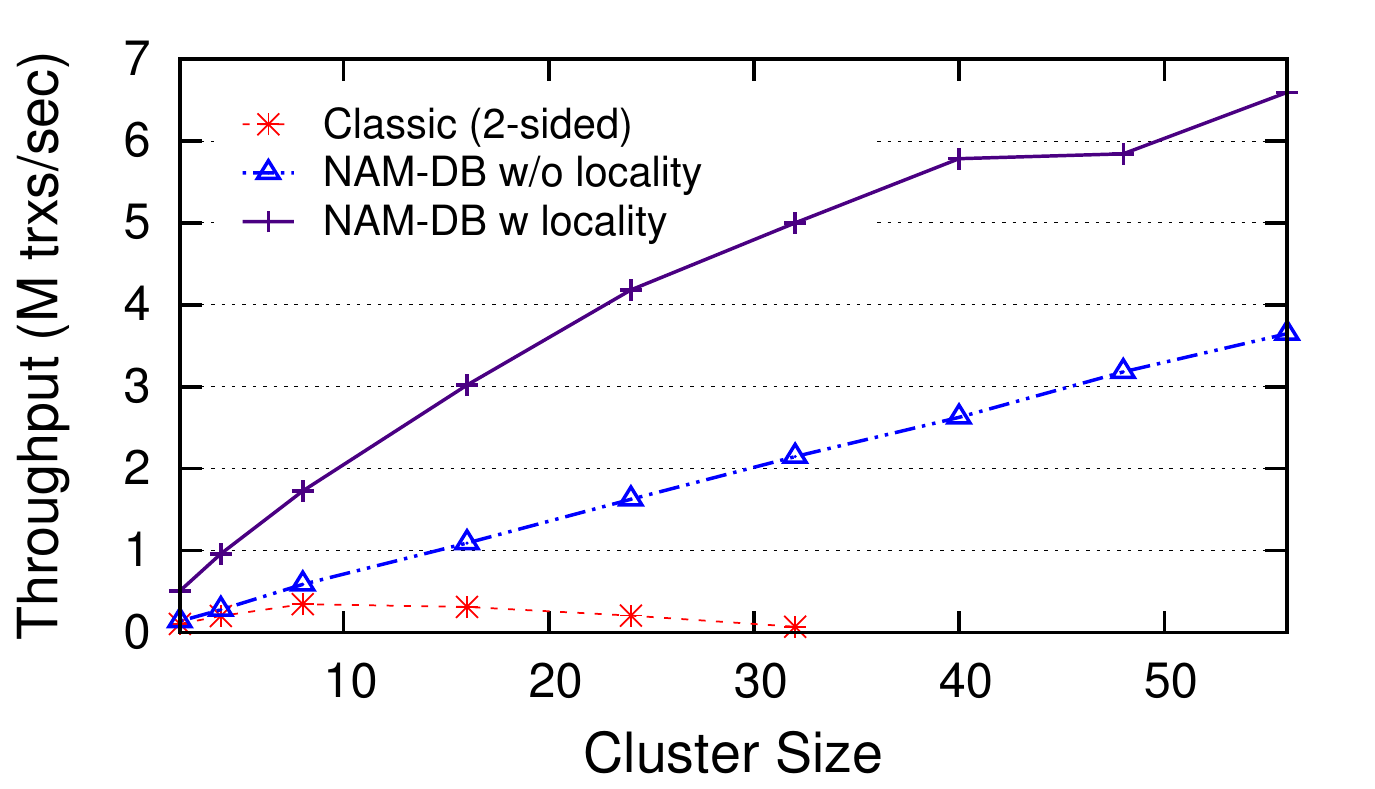}
\vspace{-3.5ex}
\caption{Scalability of \system{}}
\vspace{-4.5ex}
\label{fig:exp1}
\end{figure}

\subsection{Exp.1: System Scalability}
\label{sec:evaluation:exp1}

The main goal of this experiment is to show the scalability of \system{} for TPC-C on Cluster $A$.
We therefore increased the number of servers starting with $2$ servers and scaled-out to $56$ servers and executed \system{} with and without its locality optimization.
For the setup without locality optimization, we deployed $28$ compute servers on type 1 machines and $28$ memory servers on type 2 machines.
Each compute server uses $60$ transaction execution threads in this deployment. 
For the setup with the locality optimization, we deployed $56$ compute and $56$ memory servers (one pair per physical machine).
In this deployment, each compute server was running only $30$ transaction execution threads to have the same total number in both deployments. 
Finally,in both deployments we used one additional dedicated memory server on a type 2 machine to store the timestamp vector.
%For running the transactions, each compute server uses $60$ transaction execution threads which resulted in the highest throughput for all configurations.
%In this experiment, we used the standard configuration of TPC-C that uses $10\%$ distributed new-order transactions 

Figure \ref{fig:exp1} shows the throughput of \system{} with an increasing cluster size without exploring locality (blue), with adding locality (purple) and compares them against a more traditional implementation of Snapshot Isolation (red) with two-sided messaged based communication. 
The results show that {\bf \system{} scales nearly linear} with the number of servers to $3.64$  million distributed transactions over $56$ machines. 
This is a stunning result as in this configuration all transactions are distributed. 
However, if we allow the system to take advantage of locality we achieve even $6.5$ million TPC-C new-order transactions (recall, in the default setting roughly 10\% have to be distributed and the new-order transaction makes only up to 45\% of the workload). 
This is $2$ million more TPC-C transactions than the current scale-out record by Microsoft FaRM \cite{farm15}, which achieves $4.5$ million TPC-C transactions over $90$ machines with very comparable hardware, the same network technology and using as much locality as possible.
It should be noted though, that FaRM implements serializability guarantees, whereas \system{} supports snapshot isolation. 
While for this benchmark it makes no difference (there is no write-skew), it might be important for other workloads.
At the same time though, FaRM never tested their system for larger read queries, for which it should perform particular bad as it requires a full read-set validation. 

\begin{figure}
\centering
\subfigure[Latency]{
   \includegraphics[trim=5 0 20 0,clip,width=.46\columnwidth]{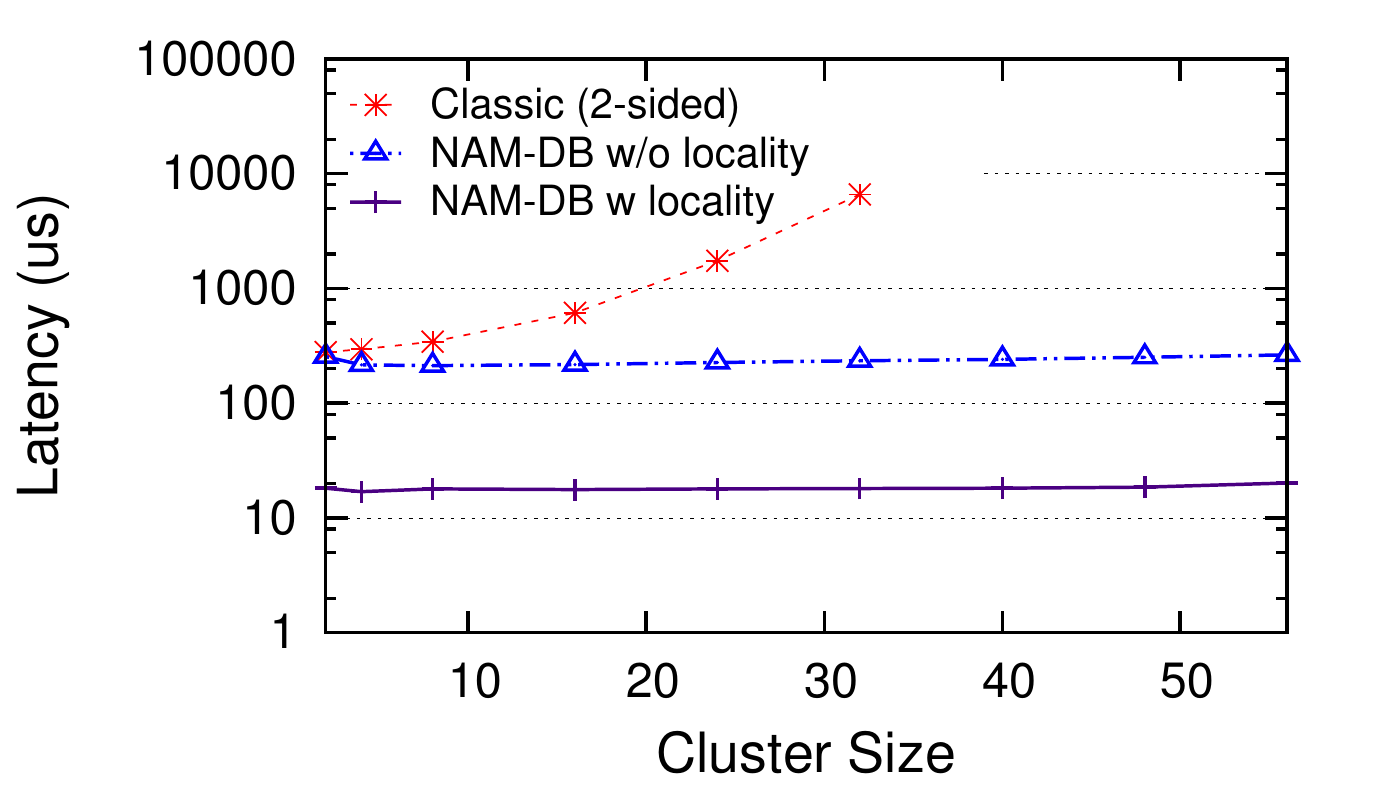}
   \label{fig:exp1b}
}
 \subfigure[Breakdown for \system{}]{
   \includegraphics[trim=5 0 20 0,clip,width=.46\columnwidth]{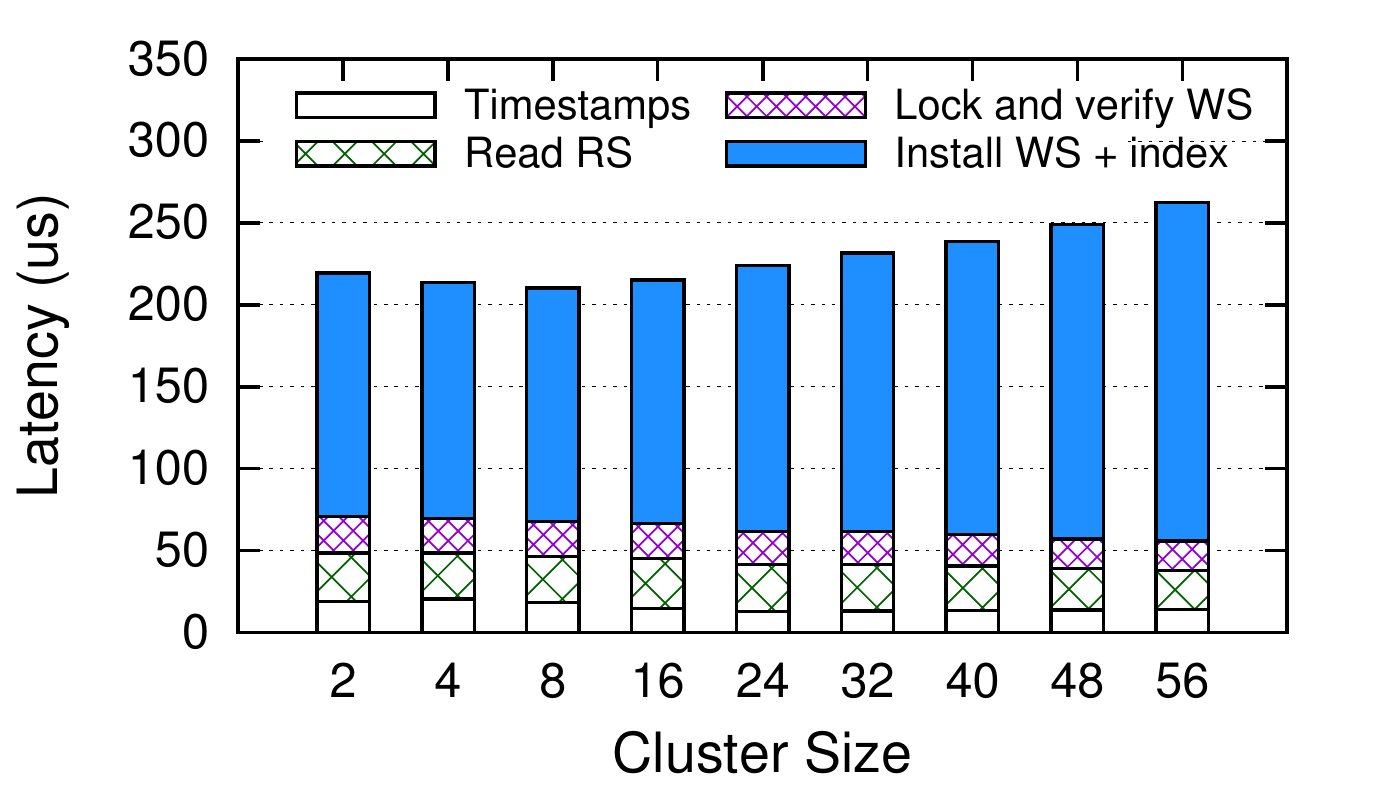}
   \label{fig:exp1c}
}
\vspace{-3.5ex}
\caption{Latency and Breakdown}
\vspace{-5.5ex}
\label{fig:exp1bc}
\end{figure}

The traditional SI protocol in Figure \ref{fig:exp1} follows a partitioned shared-nothing design similar to \cite{saphana} but using two-sided RDMA for the communication.
For this variant, we clearly see that it does not scale with the number of servers.
Even worse, the throughput even degrades when using more than $10$ machines.
This degrade results from the high CPU costs of handling messages.
This effect can also be seen by the increased latency.

Figure~\ref{fig:exp1b} shows that the latency of new-order transactions.
While \system{} almost stays constant regardless of the number of machines,  the classic SI implementation increases. 
This is not surprising as in the NAM-DB design the work per machine regardless of the number of machines in the cluster is constant, whereas the classical implementation requires more and more message handling. 

Looking more carefully into the latency of \system{} w/o locality and breaking the latency down into the different components (Figure~\ref{fig:exp1c}) reveals that the latency increases slightly mainly because of the overhead to install new version. 
Most interestingly though the latency for the timestamp oracle does not increase, indicating the efficiency of our technique (note, that we currently do not partition the timestamp vector).

Finally, another interesting question is how \system{} compares to the initial prototype in \cite{RDMAVision}.
Since the initial prototype makes many simplifying assumptions on central components (e.g., storage management), we compare \system{} when using the old and new timestamp oracle to exclude effects resulting from these simplifications.
While \system{} with the new oracle scales linearly up to 56 nodes as shown in Figure \ref{fig:exp1}, \system{} with the old oracle (not shown in the Figure \ref{fig:exp1}) provides only a linear scalability up to 10 nodes and achieves a maximum throughput of approximately 0.5 million transactions per second.
When scaling out further, the throughput however decreases to only 100.000 transactions per second on 56 nodes resulting from the contention on the old timestamp oracle; an effect that we show next.

%- Setup (Oracle): Varying servers from 1...28 (for each compute + memory servers)
%- TPC-C standard configuration 
%- Approaches:
%a) Our approach (w  and wo locality optimization)
%b) 2-sided 2PC (RDMA s/r + sockets IPoIB) 
%- 2 Plots: x=cluster size / y=throughput + latency

\subsection{Exp.2:  Scalability of the Oracle}
\label{sec:evaluation:exp4}

In this experiment, we analyze the scalability of our new timestamp oracle in isolation, which is one of the main contributions of this paper.
For testing the scalability, we vary the number of compute servers that concurrently update the oracle.
Different from the prevoius experiment, however, compute servers do not execute any real transaction logic.
Instead, each thread in a compute server executes the following actions in a closed loop to put a high load on the oracle: first it reads the current timestamp, second it generates a new commit timestamp and then it makes the new commit timestamp visible.
This sequence of operations is called a \emph{timestamp transaction} in the sequel, or simply \emph{t-trx} .
In this experiment, we use cluster $B$ with $8$ servers in total. 
Each node hosts one compute server that runs $20$ transaction execution threads whereas one node runs a memory server that stores the timestamp vector. 

\begin{figure}
\centering
\vspace{-1ex}
\includegraphics[width=0.6\columnwidth]{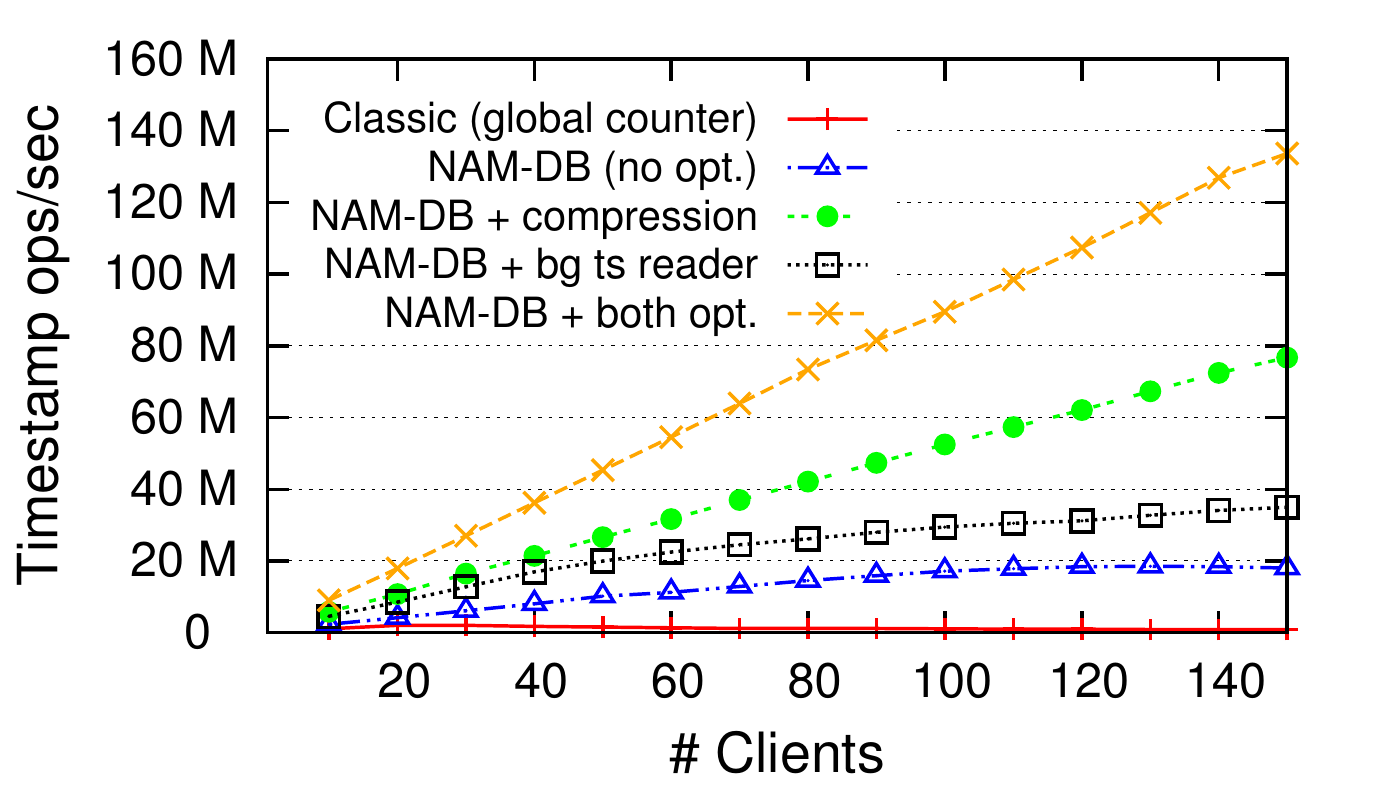}
\vspace{-3.5ex}
\caption{Scalability of Timestamp Oracle}
\vspace{-5.5ex}
\label{fig:exp4}
\end{figure}

As a baseline, we analyze the original timestamp oracle of our vision paper \cite{RDMAVision}.
This timestamp oracle implements a classical design with one globally-ordered counter, which is accessed concurrently via RDMA operations as described in Section \ref{sec:protocol}  (the red line in Figure  \ref{fig:exp4}). 
This oracle can achieve up to $2$ million t-trxs/sec. 
However, it does not scale at all with an increasing number of clients.
In fact, when scaling to more than $20$ clients, the throughput even starts to degrade due to high contention on the oracle.
However, while not scalable for large deployments, the original oracle design was still sufficient for our prototype in \cite{RDMAVision}.
The reason is that real transactions execute other operations in between reading and updating the global timestamp counter.
As a result, the maximal throughput in \cite{RDMAVision} was at most $1.1$ million TPC-W transactions per second, a load that the original oracle could sustain.
%\tim{We need to make sure, that all lines are explained in the figure. Not sure, if it is obvious right now, that line classic is the old NAM-DB design} \erfan{how about now? we mentioned in the text that this line is for the old NAM-DB, and in the figure I emphasized that it's a global counter} 

In this paper, however, the main goal is to provide a truly scalable oracle for much larger deployments. 
As presented before, when running the full mix of TPC-C transactions, our system can execute up to $6.5$ million new-order transactions on $56$ nodes.
In other words, it can execute more than $14$ million transactions (all TPC-C transactions together); a load that could not be handled by the classic oracle design (red line).

As shown in Figure \ref{fig:exp4}, we see that the new oracle (blue line) can easily sustain this load. 
For example, the basic version with no optimization achieves $20$ million t-trxs/sec. 
However, the basic version of the oracle still does not scale linearly.
The main reason is that the size of the timestamp vector grows with the number of transaction execution threads (i.e., clients) and makes the network bandwidth the bottleneck.

While $20$ million t-trxs/sec are already sufficient that the basic new oracle (blue line) does not become a bottleneck in our experiments, we can push the limit even further by applying the optimizations discussed in Section \ref{sec:oracle:opti}.
One of the optimizations is that one dedicated background fetch thread is used per compute server (instead of per transaction execution thread) to read the timestamp vector periodically in order to reduce the load on the network. 
When applying this optimization (black line, denoted by ``bg ts reader''), the oracle scales up to $36$ million t-trxs/sec.
Furthermore, when using the idea of compression (green line), where there is one entry in the timestamp vector per machine (instead of per transaction thread), the oracle scales even further to $80$ million t-trxs/sec.
Finally, when enabling both optimizations (yellow line), the oracle scales up to $135$ million t-trxs/sec on only 8 nodes.

It is worth noting that even the optimized oracle reaches its capacity at some point when deployed on clusters with hundreds or thousands of machines. At that point, the idea of partitioning the timestamp vector (see Section \ref{sec:oracle:opti}) could be applied to remove the bottleneck. Therefore, we believe that our proposed design for timestamp oracle is truly scalable.

\begin{figure}
\centering
\subfigure[Throughput]{
   \includegraphics[trim=5 0 20 0,clip,width=.46\columnwidth]{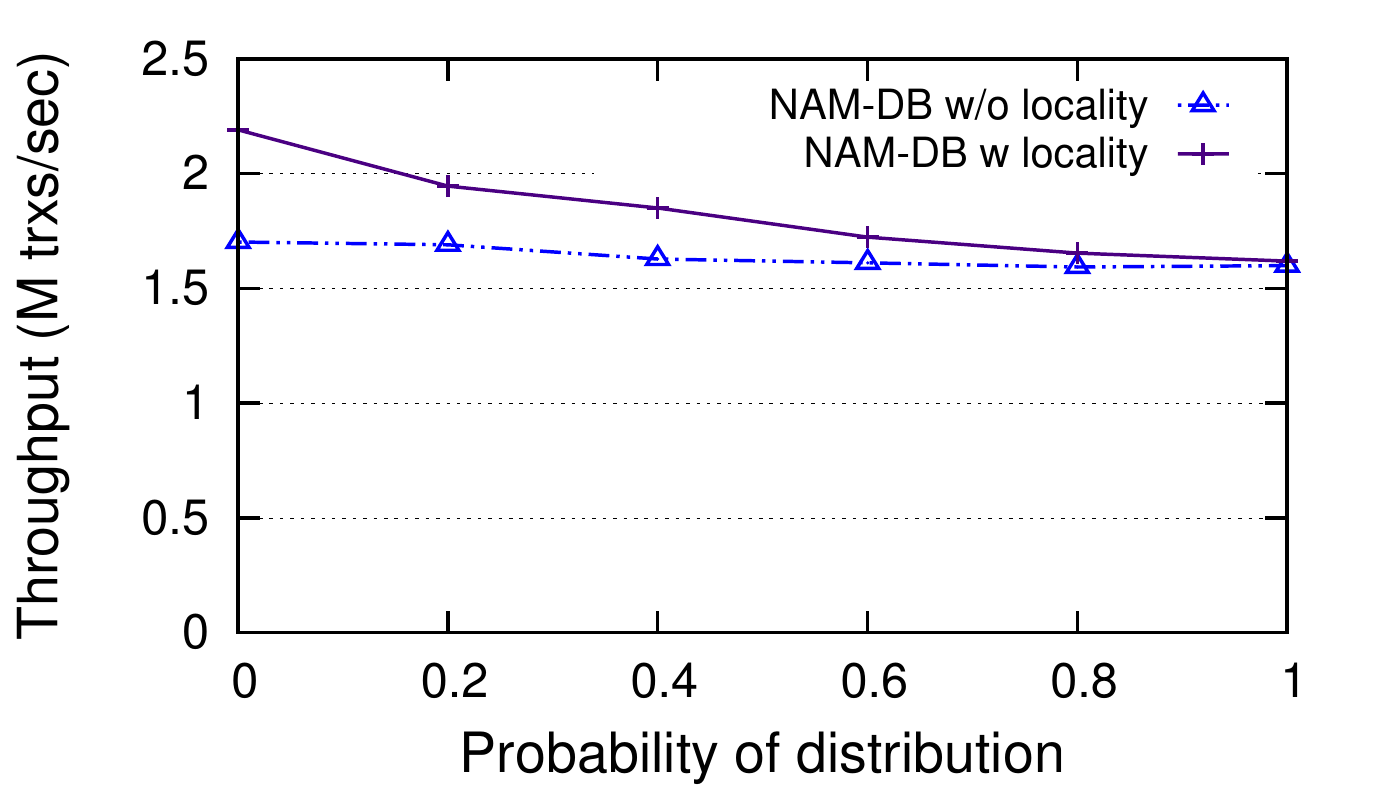}
   \label{fig:exp2a}
}
 \subfigure[Latency]{
   \includegraphics[trim=5 0 20 0,clip,width=.46\columnwidth]{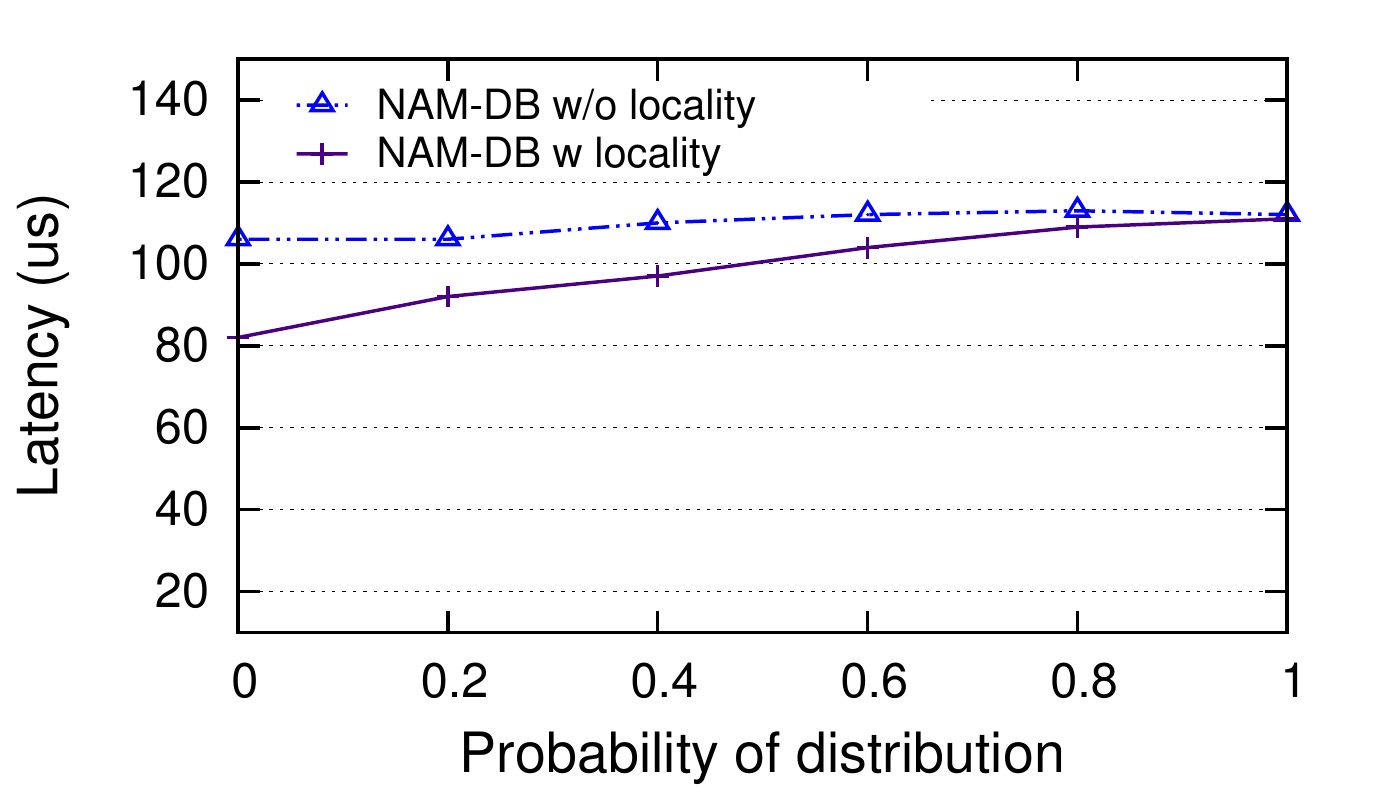}
   \label{fig:exp2b}
}
\vspace{-3.5ex}
\caption{Effect of Locality}
\vspace{-2.5ex}
\label{fig:exp2}
\end{figure}

\begin{figure}
\centering
\subfigure[Throughput]{
   \includegraphics[trim=5 0 20 0,clip,width=.45\columnwidth]{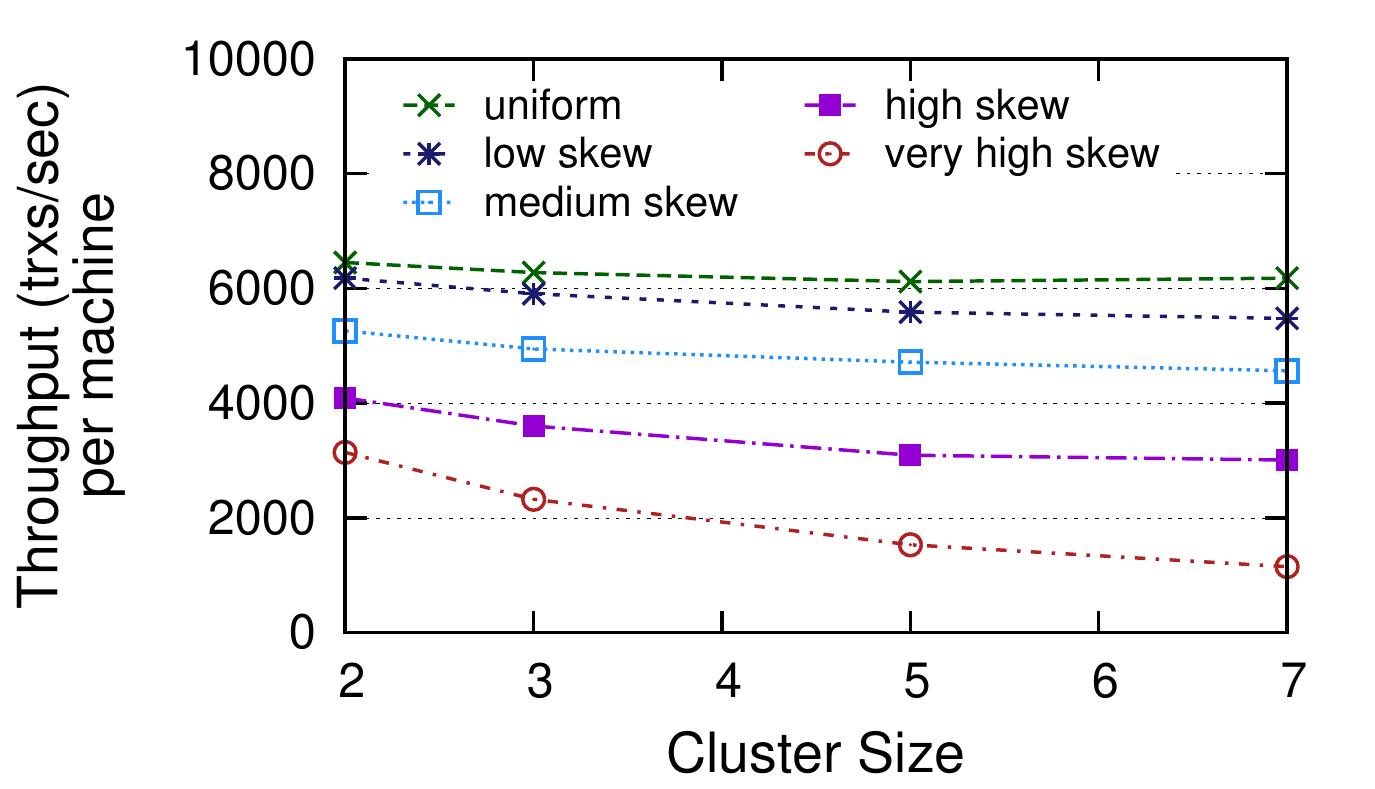}
   \label{fig:exp3a}
}
 \subfigure[Abort Rate]{
   \includegraphics[trim=5 0 20 0,clip,width=.45\columnwidth]{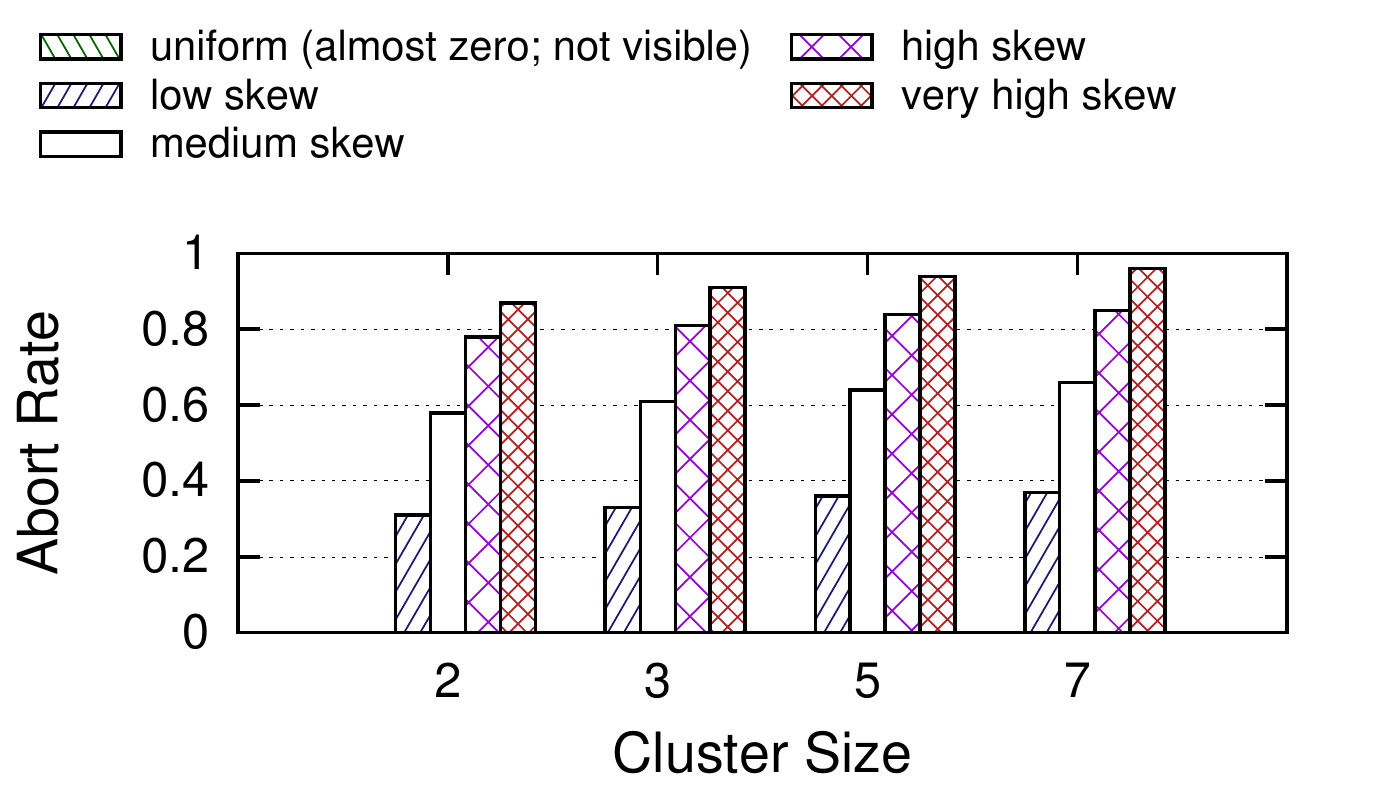}
   \label{fig:exp3b}
}
\vspace{-2.5ex}
\caption{Effect of Contention}
\vspace{-4.5ex}
\label{fig:exp3}
\end{figure}

\subsection{Exp.3: Effect of Locality}
\label{sec:evaluation:exp2}

As described earlier, we consider locality an optimization technique, like adding an index, rather than a requirement to achieve good scale-out properties. 
This is feasible as with high-speed networks the impact of locality is no longer as severe as on slow networks. 
To test this assumption, we varied the degree of distribution for new-order transactions from $0\%$ up to $100\%$.
The degree of distribution represents the likelihood that a transaction needs to read/write data from a warehouse that is stored on a remote server.
When exploiting locality, transactions are executed at those servers that store the so called home-warehouse. 
In this experiment, we only executed the new-order transaction and not the complete mix in order to show the direct effect of locality.

For the setup, we again use our cluster $B$ with $8$ servers.
One server was dedicated for the timestamp oracle.
All other $7$ machines physically co-locate one memory and one computer server.
The TPC-C database contained $200$ warehouses partitioned to all memory servers. 
When running w/o locality, we executed all memory accesses using RDMA. 
When running w/ locality we accessed directly the local memory if possible.

Figure \ref{fig:exp2} shows that the performance benefit of locality is roughly $30\%$ in regard to throughput and latency. 
While $30\%$ is not negligible it still demonstrates that there no longer are orders-of-magnitude differences between them if the system is designed to achieve high distributed transaction throughput. 

We also executed the same experiment on a modern in-memory database (H-Store \cite{Andy-thesis}) that implements a classical shared-nothing architecture which is optimized for data-locality. 
We choose H-Store as it is one of the few freely available distributed in-memory transactional database systems.
We used the distributed version of H-Store without any modifications over InfiniBand (IB) using IP over IB as communication stack.
As a main result, we found out that H-Store has an overall much lower throughput than \system{}.
On a perfectly partitionable workload (i.e., with no distributed transactions), H-Store reaches only $11$K transactions per second (not shown in Figure \ref{fig:exp2} since the line would not be visible on the given y-scale).
These numbers are in line with the ones reported in \cite{HStore:SIGMOD:2012}.
However, at $100\%$ distributed transactions the throughput of H-Store drops to only $900$ transactions per second, a $90\%$ drop in its performance, while our system still achieves more than $1.5M$ transactions under the same workload; a small performance drop compared to its no-distributed-transaction case (i.e., $0\%$ distributed transactions).
This clearly shows the sensitivity of the shared-nothing design to data locality.

%For the two classical variants that implement a shared-nothing architecture and use either two-sided RDMA operations or IPoIB for communication, we see that locality has a much stronger influence on the resulting throughput and latency of transactions than in \system{} (while the absolute throughput is still much lower even for $0\%$ distribution) .
%Both these variants are implemented to make full use of locality. 
%In order to analyze the benefits of locality, we therefore compare $0\%$ distribution and $100\%$; i.e., a degree of $100\%$ distribution means that all transactions must read/write data from remote servers while at $0\%$ all reads/writes are local.
%While for two-sided RDMA the benefits of locality result in a approx. $2\times$ higher throughput, for IPoIb the performance gain is even higher at around $5\times$ when comparing $0\%$ to $100\%$ distribution.
%The reason is that for IPoIB locality helps more since costly message passing operations of the IP stack that result in high CPU overhead can be avoided.

%- Setup (Our): Fixed number of servers (4+4)
%- Workload: TPC-C varying distribution (0...1)
%- Plot: x= degree of distr. / y=throughput + latency

\subsection{Exp.4: Effect of Contention}
\label{sec:evaluation:exp3}
In this final experiment, we analyze the effect of contention on the throughput and abort rate with an increasing number of servers and w/o locality (so all transactions are distributed) 
We vary the skew in the workload; i.e., the likelihood that a given product item is selected by a transaction by using a uniform distribution as well as different zipf distributions with \emph{low skew} ($\alpha=0.8$), \emph{medium skew} ($\alpha=0.9$), \emph{high skew} ($\alpha=1.0$) and \emph{very-high skew} ($\alpha=2.0$).
%\carsten{Erfan, please check details here!}

Figure \ref{fig:exp3} shows the results in regard to throughput and abort rate.
For the uniform and zipf distribution with low skew, we can see that the throughput per machine is stable (i.e., almost linearly as before).
However, for an increasing skewness factor the abort rate also increases due to the contention on a single machine.
This supports our initial claim that while RDMA can help us to achieve a scalable distributed database system, we can not do something against an inherently non-scalable workload that has individual contention points.
The high abort rate can be explained by the fact that we immediately about transactions instead of waiting for a lock once a transaction does not acquire a lock.
Important is that this does not have a huge impact on the throughput, since in our case the compute server directly triggers a retry after an abort.

\section{Related Work}
\label{sec:related}

In this paper, we have made the case that distributed transactions can indeed scale using the recent generation of RDMA-enabled networks technology.

Most related to our work is FaRM \cite{farm15,farm14}.
However, FaRM uses a more traditional message-based approach and focuses on serializability, whereas we implemented snapshot isolation, which is more common in practice because of its low-overhead consistent reads.
More importantly, in this work we made the case that distributed transactions can now scale, whereas FaRMs design is centered around locality.

Another recent work ~\cite{shared-database:sigmod2015} is similar to our design since it also separates storage from compute nodes. However, instead of treating RDMA as a first-class citizen, they treat RDMA as an afterthought.
Moreover, they use a centralized commit manager to coordinate distributed transactions which is likely to become a bottleneck when scaling out to larger clusters.
Conversely, our NAM-DB architecture is designed to leverage one-sided RDMA primitives to build a scalable shared distributed architecture without a central coordinator to avoid bottlenecks in the design.

Industrial-strength products have also adapted RDMA in existing DBMSs~\cite{purescale,RAC,sqlseverRDMA}.
For example, Oracle RAC~\cite{RAC} has RDMA support, including the use of RDMA atomic primitives.
However, RAC does not directly take advantage of the network for transaction processing and is essentially a work\-around for a legacy system. 
Furthermore, IBM pureScale~\cite{purescale} uses RDMA to provide high availability for DB2 but also relies on a centralized manager to coordinate distributed transactions.
Finally, SQLServer \cite{sqlseverRDMA} uses RDMA to extend the buffer pool of a single node instance but does not discuss the effect on distributed databases at all.

Other projects in aca\-demia have also recently targeted RDMA for data management, such as \emph{distributed join processing} \cite{rdmajoin,flowjoin,spinningjoin}.
However, they focus mainly only on leveraging RDMA in a traditional shared-nothing architecture and do not discuss the redesign of the full database stack.
SpinningJoins~\cite{spinningjoin} suggest a new architecture for RDMA.
Different from our work, this work assumes severely limited network bandwidth (only 1.25GB/s) and therefore streams one relation across all the nodes (similar to a block-nested loop join).
Another line of work is on \emph{RDMA-enabled key value stores}
RDMA-enabled key/value stores~\cite{ramcloud,pilaf,herd}.
We leverage some of these results to build our distributed indexes in \system{}, but transactions and query processing are not discussed in these papers. 

Furthermore, there is a huge body of work on distributed transaction processing over slow networks.
In order to reduce the network overhead, many techniques have been proposed ranging from locality-aware partitioning schemes \cite{Sword:EDBT:2013,HStore:SIGMOD:2012,Schism:VLDB:2010,LocalityAware} and speculative execution \cite{PavloWork} to new consistency levels \cite{consistencyrationing,bailis1,bailis2} and the relaxation of durability guarantees \cite{AmrWork}.

Finally, there is also recent work on high-performance OLTP systems for many-core machines \cite{foedus,hpwork,hwislands}.
While the main focus of these papers is on scaling-up on a single machine, the focus of this paper is on the aspects of scale-out when leveraging the next generation of high-speed networks and RDMA and thus can be seen as orthogonal to the other work.
%\vspace{-2.0ex}
\section{Conclusions}
\label{sec:concl}
We presented \system{}, a novel scalable distributed database system which uses distributed transactions by default and considers locality as an optimization. 
We further presented  techniques to achieve scalable timestamps for snapshot isolation as well as showed how to implement Snapshot Isolation using one-sided RDMA operations.
Our evaluation shows nearly perfect linear scale-out to up to $56$ machines and a total TPC-C throughput of $6.5$ million transactions per second, significantly more than the state-of-the-art.
In future, we plan to investigate more into avenues such as distributed index design for RDMA and to study the design of other isolation levels in more detail. 
This work ends the myth that distributed transactions do not scale and shows that \system{} is at most limited by the workload itself. 
%\vspace{-2.0ex}

\begin{scriptsize}
	
\bibliographystyle{abbrv}
\bibliography{mybib}  
\end{scriptsize}

% \clearpage
% \setcounter{figure}{0}    
% \setcounter{page}{1}
% \begin{appendices}
% \input{a_response.tex}
% \end{appendices}

\end{document}